\documentclass[useAMS,usenatbib,twocolumn]{mn2e}
\usepackage{amsmath}
\usepackage{amssymb}
\usepackage{graphicx}
\usepackage{color}
\usepackage{amsmath,amsbsy}

\newcommand{\Ma}{\mathcal{M}}

\title[]{Structure analysis of simulated molecular clouds with the $\Delta$-variance}
\author[Bertram et al.]{Erik~Bertram$^1$, Ralf~S.~Klessen$^{1,2,3}$ \& Simon~C.~O.~Glover$^1$\\
$^1$Universit\"at Heidelberg, Zentrum f\"ur Astronomie, Institut f\"ur Theoretische  Astrophysik, Albert-Ueberle-Str.~2, 69120 Heidelberg, Germany \\
$^2$Department of Astronomy and Astrophysics, University of California, 1156 High Street, Santa Cruz, CA 95064, USA \\
$^3$Kavli Institute for Particle Astrophysics and Cosmology, Stanford University, SLAC, Menlo Park, CA 94025, USA
}

\begin{document}

\maketitle

\abstract
We employ the $\Delta$-variance analysis and study the turbulent gas dynamics of simulated molecular clouds (MCs). Our models account for a simplified treatment of time-dependent chemistry and the non-isothermal nature of the gas. We investigate simulations using three different initial mean number densities of $n_{0} = 30, 100$ and $300 \: {\rm cm^{-3}}$ that span the range of values typical for MCs in the solar neighbourhood. Furthermore, we model the CO line emission in a post-processing step using a radiative transfer code. We evaluate $\Delta$-variance spectra for centroid velocity (CV) maps as well as for integrated intensity and column density maps for various chemical components: the total, H$_2$ and $^{12}$CO number density and the integrated intensity of both the $^{12}$CO and $^{13}$CO ($J=1 \rightarrow 0$) lines. The spectral slopes of the $\Delta$-variance computed on the CV maps for the total and H$_2$ number density are significantly steeper compared to the different CO tracers. We find slopes for the linewidth-size relation ranging from 0.4 to 0.7 for the total and H$_2$ density models, while the slopes for the various CO tracers range from 0.2 to 0.4 and underestimate the values for the total and H$_2$ density by a factor of $1.5-3.0$. We demonstrate that optical depth effects can significantly alter the $\Delta$-variance spectra. Furthermore, we report a critical density threshold of $\sim100\,$cm$^{-3}$ at which the $\Delta$-variance slopes of the various CO tracers change sign. We thus conclude that carbon monoxide traces the total cloud structure well only if the average cloud density lies above this limit.
\endabstract


\section{Introduction}
\label{sec:introduction}

The interstellar medium (ISM) is dominated by highly turbulent motions, which contribute to regulating stellar birth in molecular clouds (MCs) \citep[see, e.g.][]{MacLowAndKlessen2004,ScaloAndElmegreen2004,ElmegreenAndScalo2004,McKeeAndOstriker2007,HennebelleAndFalgarone2012}. On large scales, turbulent gas motions support MCs against gravitational collapse. However, the shocks associated with supersonic turbulence will create overdense regions on small scales, which in turn may collapse and form stars and clusters. In addition, ISM turbulence influences other physical processes such as the chemical makeup of the gas and the associated heating and cooling processes, the efficiency with which the external radiation field will be able to penetrate into the interior of dense clouds, as well as the overall magnetic field. Better understanding the role of turbulence in the ISM, therefore, is of pivotal importance for many fields of modern astrophysics \citep[see also the lecture notes by][]{KlessenAndGlover2014}.

Comparing theoretical simulations to observational measurements is very difficult, since observations are always a complex convolution of the density and the velocity field, affected by several other important physical processes (e.g. by magnetic fields, the chemistry, stellar feedback, etc.). Furthermore, it is often difficult to infer reliable physical parameters from observational measurements. For example, observers rely on abundant tracers, e.g. $^{12}$CO, in order to measure the amount of H$_2$ gas in MCs. An obstacle with the line emission of $^{12}$CO is that it becomes optically thick in dense cloud regions and thus is not a good tracer for those regions anymore. To avoid this problem, observers also employ other tracers to study the mass distribution within a cloud, e.g. dust or $^{13}$CO, which is often optically thin. In addition, molecular tracers like CS, HCN, HCO$^+$ or NH$_3$ can be used to study the gas mass in very dense cloud regions. However, comparisons with simulations are needed in order to study the influence of chemical inhomogeneities and optical depth effects on observational measurements.

\citet{LisEtAl1996} introduced maps of centroid velocities (CV) and showed that those are to a certain degree sensitive to the underlying physics. Accordingly, it is possible to use two-point statistics in order to recover important information (e.g. optical depth effects, density and velocity fluctuations, etc.) of an astrophysical system, e.g. by using centroid velocity increment structure functions \citep[see, e.g.][]{Hily-BlantEtAl2008,FederrathEtAl2010}, spectral correlation functions \citep{RosolowskyEtAl1999}, the velocity channel analysis \citep{LazarianEtAl2000,LazarianEtAl2004}, the Principal Component Analysis \citep{HeyerAndSchloerb1997,BruntAndHeyer2002a} or the $\Delta$-variance \citep[see, e.g.][]{StutzkiEtAl1998,MacLowAndOssenkopf2000,OssenkopfEtAl2001,BenschEtAl2001,OssenkopfEtAl2008a,OssenkopfEtAl2008b}. The latter provides a wavelet-based measure for characterizing structures in astronomical datasets. For example, \citet{StutzkiEtAl1998} used the $\Delta$-variance in order to measure the scaling behavior of structures in observed images by analyzing the line emission of $^{12}$CO and $^{13}$CO, while \citet{BenschEtAl2001} studied the influence of white noise and beam smoothing on the $\Delta$-variance spectra. Moreover, \citet{OssenkopfEtAl2008b} applied the $\Delta$-variance to interstellar turbulence and observations of MCs and tested the capabilities of the method in a practical use by applying different filter functions with different diameter ratios.

Several studies tried to reveal the influence of projection effects on different statistical quantities, finding that various physical processes might influence the projection of three-dimensional data on a two-dimensional map on the sky. For example, \citet{LazarianEtAl2004}, \citeauthor{BurkhartEtAl2013b}~(2013b), \citet{BurkhartEtAl2014} and \citeauthor{BertramEtAl2015a}~(2015a) have shown that optical depth effects might significantly alter the statistics of centroid velocities due to different opacities of the gas tracers. Furthermore, the turbulent driving as well as density and temperature fluctuations along the line-of-sight also have a significant impact on the CV statistics \citep{LazarianAndEsquivel2003,OssenkopfEtAl2006,EsquivelEtAl2007,Hily-BlantEtAl2008}. Much progress has been made during the last years in this context, although we are still missing a consistent picture of turbulence theory in the ISM.

In this paper we study high-resolution, 3D and time-dependent chemistry models of hydrodynamical numerical simulations of the turbulent ISM. We use the $\Delta$-variance method \citep[see, e.g.][]{StutzkiEtAl1998,BenschEtAl2001,OssenkopfEtAl2008a} in order to study the structure of MCs and analyze how different chemical tracers affect the $\Delta$-variance spectra. We also vary the initial number density in order to explore how different densities affect the statistics. Moreover, we perform a radiative transfer post-processing and produce synthetic maps as well as position-position-velocity (PPV) cubes of the $^{12}$CO and $^{13}$CO emission. We then compute maps of centroid velocities, column densities and integrated intensities in order to analyze how the chemical inhomogeneity and the variable opacity affect the various $\Delta$-variance spectra.

In Section \ref{sec:methodsandsims} we present our numerical simulations, introduce the radiative transfer post-processing and explain the statistical methods. In Section \ref{sec:results} we present our results, which are discussed in Section \ref{sec:discussion}. Finally, we summarize our findings and present our conclusions in Section \ref{sec:summary}.

\section{Methods and simulations}
\label{sec:methodsandsims}

The simulation data presented in the following sections are also used in \citet{BertramEtAl2014} and \citeauthor{BertramEtAl2015a}~(2015a). While \citet{BertramEtAl2014} use the technique of Principal Component Analysis (PCA) in order to statistically analyze the turbulent flows in the spectral data cubes, \citeauthor{BertramEtAl2015a}~(2015a) evaluate the slopes of the structure functions of 2D projected centroid velocities and compute Fourier spectra. Here, we summarize the most important aspects of our hydrodynamical simulations and the radiative transfer post-processing.

\subsection{Computational method}
\label{subsec:sims}

The simulations in this paper were performed using a modified version of the Z{\sc eus}-M{\sc p} MHD code \citep{Norman2000,HayesEtAl2006}. We have embedded a detailed atomic and molecular cooling function, described in \citet{GloverEtAl2010} and \citet{GloverAndClark2012}, together with a simplified treatment of the molecular gas chemistry. The chemical network is based on the work of \citet{NelsonAndLanger1999} and \citet{GloverAndMacLow2007}, and allows us to follow the formation and destruction of H$_{2}$ and CO self-consistently within our simulations. The network tracks the abundances of 9 species and follows 30 chemical reactions. We adopt the standard solar abundances of hydrogen and helium. The abundances of carbon and oxygen are taken from \citet{SembachEtAl2000}. We use $x_{\text{C}} = 1.4 \times 10^{-4}$ and $x_{\text{O}} = 3.2 \times 10^{-4}$, where $x_{\text{C}}$ and $x_{\text{O}}$ are the fractional abundances by number of carbon and oxygen relative to hydrogen \citep[for further reading see also][]{GloverEtAl2010,GloverAndMacLow2011}. When we start our simulations, the carbon is assumed to be singly ionised as C$^{+}$, while the hydrogen, helium as well as the oxygen are in atomic form. For further information about the chemical model we refer the reader to \citet{GloverAndClark2012}.

We run simulations with three different initial number densities $n_{0}$ of the hydrogen nuclei, $n_{0} = 300, 100$ and $30 \: {\rm cm^{-3}}$. The temperature of the gas at the beginning of our runs is set to a constant value of 60\,K. We use an uniform weak magnetic field strength $B_{0} = 5.85 \mbox{ }\mu {\rm G}$, which is initially oriented parallel to the $z$-axis of the computational domain. We do not include self-gravity. The gas is initially uniform and embedded in a periodic box with a side length of 20\,pc. The turbulent simulations are uniformly driven between wavenumbers $k = 1$ and $k = 2$ \citep{MacLowKlessenBurkertSmith1998,MacLow1999} with a 3D rms velocity of $v_{\rm rms} = 5 \: {\rm km \: s^{-1}}$. This value remains approximately constant throughout the whole simulation period. For the dust-to-gas ratio, we adopt the standard local value of 1:100 \citep{GloverEtAl2010}, and assume that the dust properties do not vary with the gas density. The cosmic ray ionization rate was set to $\zeta = 10^{-17} \: {\rm s^{-1}}$. For the incident ultraviolet radiation field, we adopt the standard parameterization of \citet{Draine1978}. This field has a strength $G_0 = 1.7$ in \citet{Habing1968} units, corresponding to an integrated flux of $2.7 \times 10^{-3}\,$erg\,cm$^{-2}$s$^{-1}$. We use a numerical resolution of $512^3$ grid cells and discuss the influence of our limited numerical resolution in more detail in Appendix \ref{app:resolution}. Using this value for the resolution as well as the scale of the total box, we can compute the cubic cell size and obtain $\Delta x \approx 0.04\,$pc for our numerical simulations.

\subsection{Radiative transfer post-processing}
\label{subsec:RADMC}

We use the radiative transfer code R{\sc admc}-3{\sc d}\footnote{www.ita.uni-heidelberg.de/$\sim$dullemond/software/radmc-3d/} \citep{Dullemond2012} in order to model the CO ($J=1 \rightarrow 0$) line for both $^{12}$CO and $^{13}$CO. Furthermore, we apply the Large Velocity Gradient (LVG) approximation \citep{Sobolev1957} to compute the level populations, as explained by \citeauthor{ShettyEtAl2011a}~(2011a). Beyond the line emission, R{\sc admc}-3{\sc d} also accounts for the dust continuum emission, which we subtract off before we analyze the statistical properties of the synthetic maps. We use a number of 512 channels in velocity space for our radiative transfer post-processing, corresponding to a spectral resolution of $\sim0.07\,$km\,s$^{-1}$, $\sim0.06\,$km\,s$^{-1}$ and $\sim0.05\,$km\,s$^{-1}$ for the n300, n100 and n30 model, respectively.

Our simulations do not explicitly track the abundance of $^{13}$CO (which would be costly), and so we need a procedure to relate the $^{13}$CO number density to that of $^{12}$CO. A common assumption is that the ratio of $^{12}$CO to $^{13}$CO is identical to the elemental abundance ratio of $^{12}$C to $^{13}$C \citep[see, e.g.][]{RomanDuvalEtAl2010}. In the majority of our analysis, we make the same assumption and set the $^{12}$CO to $^{13}$CO ratio to a constant value, $R_{12/13} = 50$, which we use to generate a first set of $^{13}$CO column density maps. However, the effects of chemical fractionation \citep{WatsonEtAl1976} and selective photodissociation of $^{13}$CO \citep[see, e.g.][]{VisserEtAl2009} can significantly alter the value of $R_{12/13}$ within the cloud \citep{RoelligAndOssenkopf2013,SzucsEtAl2014}. Therefore, we also explore the effect that this may have on the statistics of the $\Delta$-variance, using numerical results of \citet{SzucsEtAl2014}. \citet{SzucsEtAl2014} give a numerical fitting formula, which relates the ratio $R_{12/13}$ to the $^{12}$CO column density in the cloud. Thus, we also produce a second set of maps using their routine and compute variable $^{13}$CO number densities by dividing the $^{12}$CO number densities by the mean value of $R_{12/13}$ from the fitting formula for each line-of-sight. Finally, we can use these $^{13}$CO number densities in order to compute the $^{13}$CO emission in the same fashion as in our constant ratio models. The differences in the $\Delta$-variance spectra of the two sets of $^{13}$CO emission maps - those derived using a constant $R_{12/13}$ and those that use a spatially varying value of $R_{12/13}$ - will be discussed in Section \ref{subsec:ab13CO} below.

The radiative transfer calculation yields position-position-velocity (PPV) cubes of brightness temperatures $T_B$, which are related to the intensity via the Rayleigh-Jeans approximation,
\begin{equation}
\label{eq:RQ}
T_B(\nu) = \left ( \frac{c}{\nu} \right )^2 \frac{I_{\nu}}{2k_B},
\end{equation}
where $I_{\nu}$ is the specific intensity at frequency $\nu$ and $k_B$ the Boltzmann constant. We will refer to these as the ``intensity'' models. In analogy, we construct centroid velocity and column density maps out of the PPP simulation data of the density and velocity field, as described in the following Section \ref{subsec:CV}. We will refer to these as the ``density'' models, following the notation already used in \citet{BertramEtAl2014} and \citeauthor{BertramEtAl2015a}~(2015a).

\subsection{Centroid velocity, integrated intensity and column density maps}
\label{subsec:CV}

One part of our study is based on the statistics of centroid velocities. The centroids $C(x, y) = C(\textbf{r})$ are defined as
\begin{equation}
\label{eq:CV}
C(\textbf{r}) = \frac{\int F(\textbf{r}, z) v_z(\textbf{r}, z)\text{d}z}{\int F(\textbf{r}, z)\text{d}z},
\end{equation}
where the variable $v_z(\textbf{r}, z)$ is the velocity component along the line-of-sight (e.g. the $z$-direction). Thus, $C(\textbf{r})$ is a map of line-of-sight projected velocities. The quantity $F(\textbf{r}, z)$ is a statistical weight. It denotes either the underlying density field or the brightness temperatures from the PPV cubes. As shown by \citet{LisEtAl1996}, centroid velocity maps are very sensitive to the underlying physics of a MC, which means that they can be used in order to infer important information about the astrophysical system. Furthermore, CV maps can be easily computed from observational data that rely on spectral measurements. These can be then translated into maps of centroid velocities via equation (\ref{eq:CV}).

Beside the CV maps, we also consider maps of integrated intensity and column density. We calculate the former along a given line-of-sight (e.g. the $z$-direction) via
\begin{equation}
\label{eq:W}
W(\textbf{r}) = \int T_B(\textbf{r}, z) \, \text{d}v,
\end{equation}
where $T_B$ is the brightness temperature, as introduced in Section \ref{subsec:RADMC}. The latter is given by
\begin{equation}
\label{eq:N}
N(\textbf{r}) = \int n(\textbf{r}, z) \, \text{d}z,
\end{equation}
where $n$ is the number density of an individual chemical component in a given cell in the simulation box.

\subsection{The $\Delta$-variance method}
\label{subsec:DV}

\begin{figure}
\centerline{
\includegraphics[width=1.0\linewidth]{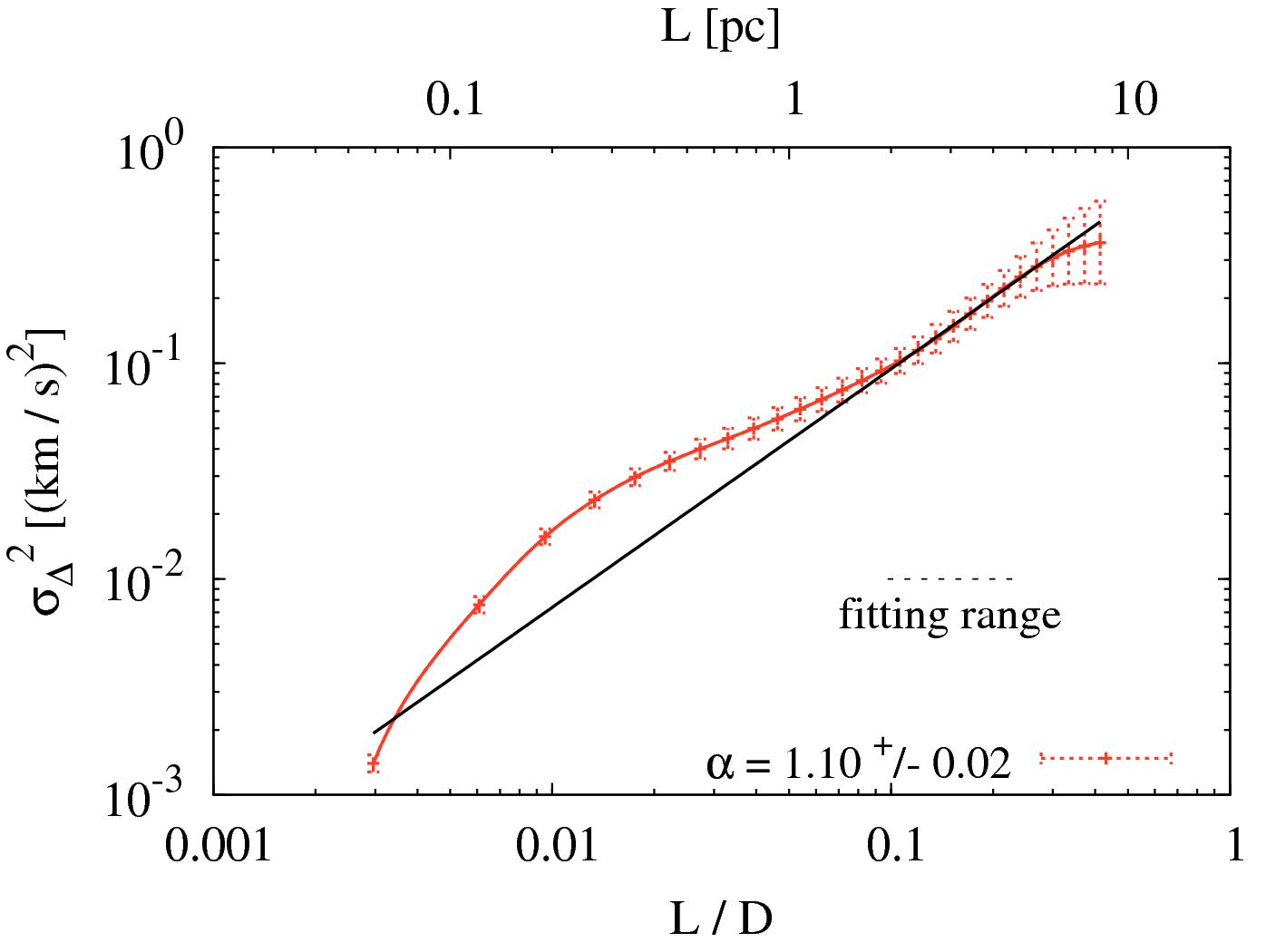} \\
}
\caption{Example of a $\Delta$-variance spectrum, plotted as a function of spatial scale, normalized by the total box size and averaged over all available time snapshots. The inferred slope $\alpha$ is also indicated (solid line). The $\Delta$-variance is computed on a CV map, where the velocities are weighted by the total density field for an initial number density of $n_{0} = 300$~cm$^{-3}$. The fitting range is indicated by a horizontal dashed line. Error bars denote 1-$\sigma$ spatial and temporal variations.}
\label{fig:DV}
\end{figure}

We use the $\Delta$-variance tool developed by \citet{OssenkopfEtAl2008a}\footnote{http://hera.ph1.uni-koeln.de/~ossk/Myself/deltavariance.html}. The $\Delta$-variance method measures the variance in a structure $S(\textbf{r})$ (in our case the maps of centroid velocities, integrated intensities and column densities) on a given spatial scale $\ell$, by filtering the dataset with a spherically symmetric up-down-function of size $\ell$. It is given by
\begin{equation}
\label{eq:DV}
\sigma_\Delta^2 (\ell) = \Bigl\langle \Big( S(\textbf{r}) \ast \bigodot_{\ell} (\textbf{r}) \Big)^2 \Bigl\rangle_{\textbf{r}},
\end{equation}
where the average is computed over all positions $\textbf{r} = (x, y)$ on the sky. The symbol $\ast$ stands for a convolution and $\bigodot_{\ell}$ describes the filter function. In this paper, we use a Mexican hat with a diameter ratio of 1.5. However, we have also analyzed the impact of the filter function and the diameter ratio on the slope values of the spectra, e.g. by using a French hat with a diameter ratio of 3.0. We find that our results do not significantly depend on the specific choice of the filter function and the diameter ratio (see Appendix \ref{app:filters}). The differences between the individual filter functions are described in more detail in \citet{OssenkopfEtAl2008a}.

In this study we compute the $\Delta$-variance for each possible line-of-sight direction $x$, $y$ and $z$. As shown by \citet{EsquivelAndLazarian2005} and \citet{BurkhartEtAl2014}, the statistics of velocity centroids are very sensitive to the direction of the magnetic field in the regime of sub-Alfv\'enic Mach numbers. Although we use a weak magnetic field in the $z$-direction, the turbulence in our simulations is trans-Alfv\'enic or mildly super-Alfv\'enic (see Table \ref{tab:setup}). The field lines are essentially dragged along with the turbulent flow, with the result that the turbulence remains approximately isotropic. Hence, we do not find significant variations of the $\Delta$-variances along the different directions. We therefore average all $\Delta$-variances of the three line-of-sights. Finally, power-laws of the form
\begin{equation}
\sigma_\Delta^2 (\ell) \propto \ell^{\alpha}
\end{equation}
were fit to the $\Delta$-variance spectra, where $\alpha$ denotes the slope of the power-law. To calculate the scaling exponents, we use a fitting range from $1/10$ to $1/4$ of the total box size ($0.1 \lesssim \ell/D \lesssim 0.25$), as constrained by \citet{FederrathEtAl2010} and \citet{KonstandinEtAl2012a}. For a box with $D = 512$ grid cells for each side this translates to 51 and 128 cells, corresponding to a physical scale of $\sim2-5\,$pc in the simulation domain. Extending the fitting range to scales above this limit is complicated since the simulations are driven on large scales, which would significantly bias our results. Additionally, the scales below our limit are influenced by the numerical resolution and the bottleneck effect \citep{KritsukEtAl2007,KonstandinEtAl2015}, which is an accumulation of kinetic energy caused by the viscosity of the fluid before it dissipates into thermal energy. Furthermore, as demonstrated by \citet{KonstandinEtAl2015}, we also note that a clear and unique identification of the fitting range in numerical simulations with our resolution is complicated. Thus, the specific choice of the fitting range introduces some degree of uncertainty in the $\Delta$-variance slopes, which we conservatively estimate to be of the order of $\pm 0.1$.

Fig. \ref{fig:DV} shows an example of a $\Delta$-variance spectrum computed using maps of centroid velocities inferred from our n300 run. The velocities are weighted by the total density field and the spectrum is averaged over all available time snapshots. We also show the fitting slope $\alpha$ for this spectrum. This slope can be translated into a linewidth-size relation, given by $\sigma_\Delta (\ell) \propto \ell^{\gamma}$, with a scaling exponent $\gamma = \alpha/2$. The latter equation relates the velocity fluctuations to their spatial scale, which is often referred to as Larson's law \citep{Larson1981}. This relation suggests that turbulence plays an important role in the process of star formation and predicts a turbulent energy cascade within the cloud, as proposed by \citet{Kolmogorov1941}. Hence, the slopes $\gamma$ characterize the turbulent velocity hierarchy in our clouds and can be directly compared to values derived from spectral observations of MCs, typically ranging from $\gamma \approx 0.2-0.9$ \citep[see, e.g.][]{Larson1981,SolomonEtAl1987,BruntEtAl2002,HeyerAndBrunt2004,HeyerEtAl2006,SunEtAl2006,Hily-BlantEtAl2008,RomanDuvalEtAl2011,SchneiderEtAl2011,RusseilEtAl2013,EliaEtAl2014,AlvesdeOliveiraEtAl2014,JohnstonEtAl2014}. For further reading about the theory of turbulence in astrophysics and fluids, we refer the reader to \citet{Burgers1948}, \citet{BenziEtAl1993}, \citet{Frisch1995}, \citet{SheAndLeveque1994}, \citet{Dubrulle1994}, \citet{BoldyrevEtAl2002} or \citet{MacLowAndKlessen2004}.

Furthermore, the scaling exponents $\alpha$ of the $\Delta$-variance spectra are related to the power-law exponents $\beta$ of the corresponding Fourier spectra via $\alpha = \beta - 1$ for 3D data \citep{StutzkiEtAl1998}, where the power-law spectrum in Fourier space is defined as
\begin{equation}
\label{eq:PS}
P(k) \propto k^{-\beta}.
\end{equation}
In this equation, $k = 2\pi / \ell$ denotes the wavevector. In analogy, for a power spectrum of a 2D image, the $\Delta$-variance is related to the power spectrum via $\alpha = \beta - 2$.

We also note that there are many other techniques to measure structural density and velocity fluctuations as a function of spatial scale, e.g. by computing structure functions \citep[see, e.g.][]{Hily-BlantEtAl2008,FederrathEtAl2010} or the spectral correlation function \citep{RosolowskyEtAl1999}, by using the velocity channel analysis \citep{LazarianEtAl2000,LazarianEtAl2004} or the Principal Component Analysis \citep{HeyerAndSchloerb1997,BruntAndHeyer2002a,BertramEtAl2014}.

\section{Results}
\label{sec:results}

\begin{table}
\begin{tabular}{l|c|c|c}
\hline\hline
Model name & n300 & n100 & n30 \\
\hline
Mean density $\lbrack$cm$^{-3}\rbrack$ & 300 & 100 & 30 \\
Resolution & 512$^3$ & 512$^3$ & 512$^3$ \\
Box size [pc] & 20 & 20 & 20 \\
$t_{\text{end}}$ [Myr] & 5.7 & 5.7 & 5.7 \\
$\langle \Ma_{\text{s}} \rangle$ & 10.6 & 6.8 & 5.1 \\
$\langle \Ma_{\text{A}} \rangle$ & 1.5 & 1.1 & 1.0 \\
$\sigma_{\rho} / \langle \rho \rangle$ & 3.0 & 3.0 & 4.6 \\
$\langle x_{\text{H}_2} \rangle_{\text{mass}}$ & 0.98 & 0.78 & 0.61 \\
$\langle n_{\text{H}_2} \rangle_{\text{vol}}$ $\lbrack$cm$^{-3}\rbrack$ & 140 & 37 & 9 \\
$\langle n_{\text{H}_2} \rangle_{\text{mass}}$ $\lbrack$cm$^{-3}\rbrack$ & 1456 & 447 & 264 \\
$\langle n_{\text{CO}} \rangle_{\text{vol}}$ $\lbrack$cm$^{-3}\rbrack$ & $2.2 \times 10^{-2}$ & $1.7 \times 10^{-3}$ & $1.4 \times 10^{-4}$\\
$\langle n_{\text{CO}} \rangle_{\text{mass}}$ $\lbrack$cm$^{-3}\rbrack$ & 0.3 & $4.3 \times 10^{-2}$ & $2.3 \times 10^{-2}$ \\
$\langle T \rangle_{\text{vol}}$ $\lbrack$K$\rbrack$ & 35 & 68 & 223 \\
$\langle T \rangle_{\text{mass}}$ $\lbrack$K$\rbrack$ & 13 & 26 & 57 \\
$\langle N_{\text{H}_2} \rangle$ $\lbrack$cm$^{-2}\rbrack$ & $8.6 \times 10^{21}$ & $2.3 \times 10^{21}$ & $5.3 \times 10^{20}$\\
$\langle N_{\text{CO}} \rangle$ $\lbrack$cm$^{-2}\rbrack$ & $1.3 \times 10^{18}$ & $1.1 \times 10^{17}$ & $8.7 \times 10^{15}$ \\
\hline
\end{tabular}
\caption{Overview of our different models with some characteristic values measured for the last time snapshot. From top to bottom we list: mean number density, resolution, box size, time of the last snapshot, mean sonic Mach number, mean Alfv\'en Mach number, ratio of density dispersion and mean density, mass-weighted mean abundances of H$_2$ \citep[i.e. the percentage of atomic hydrogen that has been converted to H$_2$, see][]{GloverAndMacLow2011}, mean volume- and mass-weighted H$_2$ and CO number densities, mean volume- and mass-weighted temperature and mean column densities of H$_2$ and CO.}
\label{tab:setup}
\end{table}

\begin{figure*}
\centerline{
\includegraphics[height=0.26\linewidth]{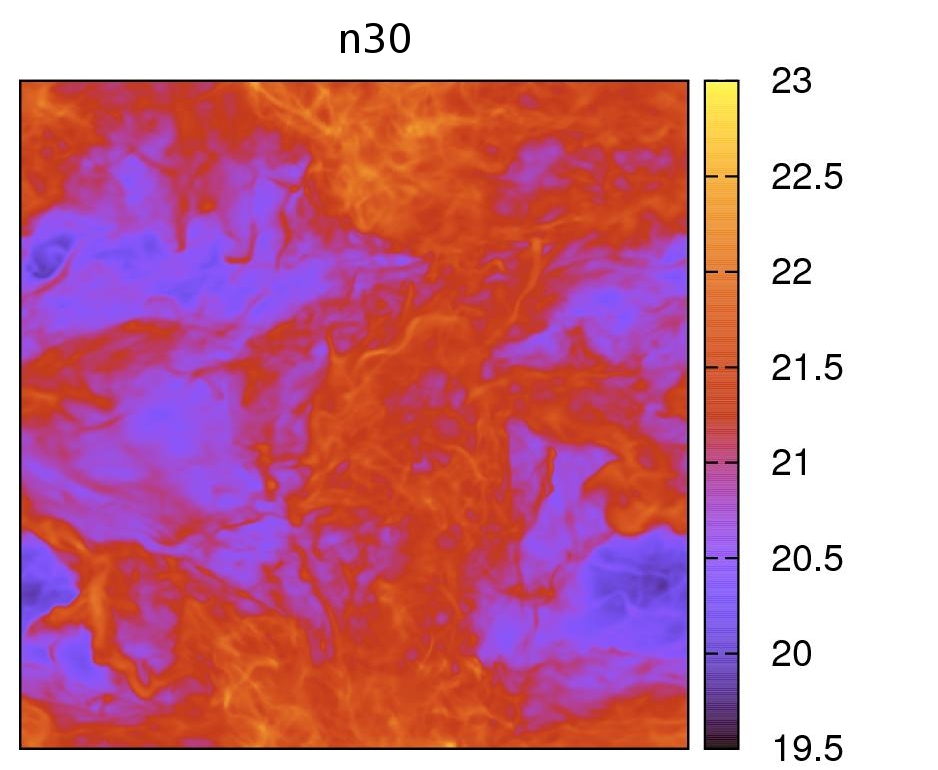}
\includegraphics[height=0.26\linewidth]{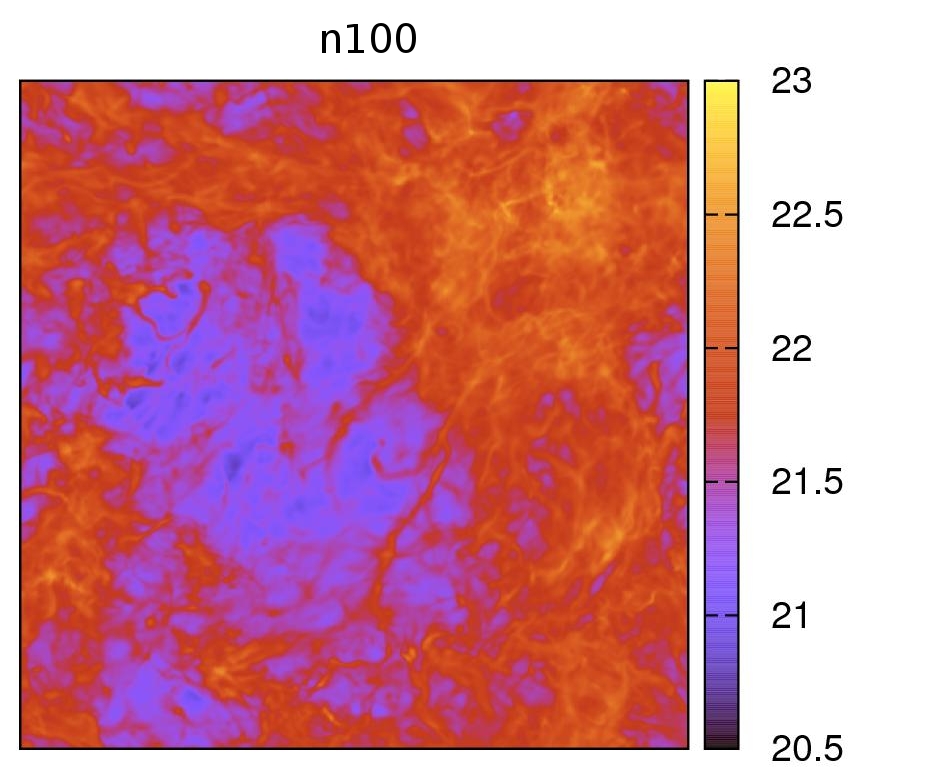}
\includegraphics[height=0.26\linewidth]{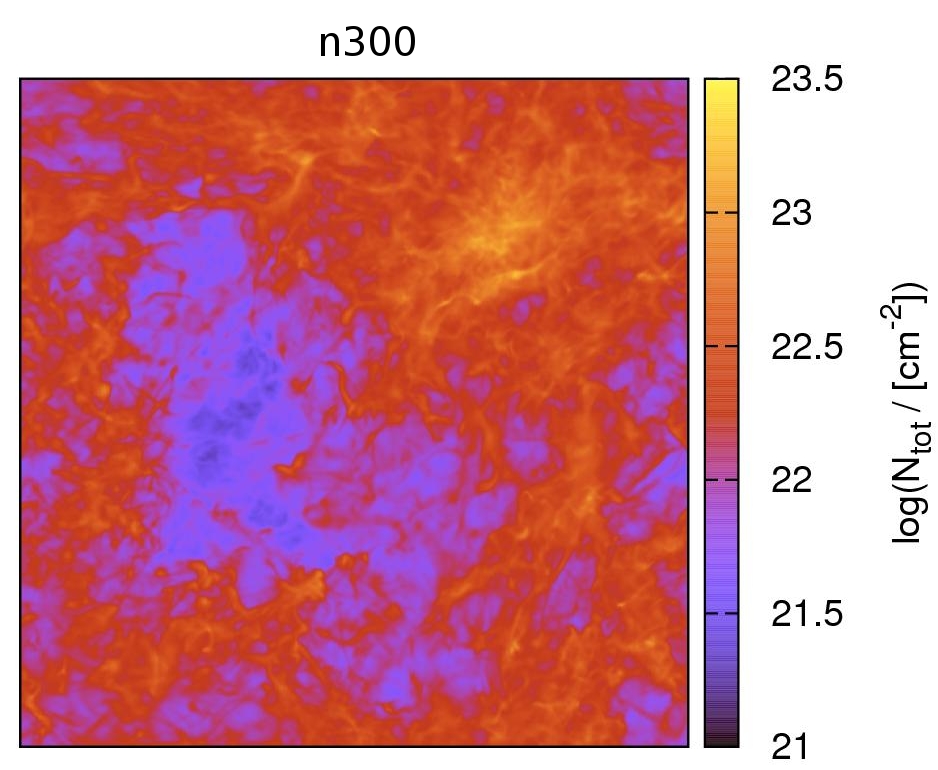}
}
\centerline{
\includegraphics[height=0.24\linewidth]{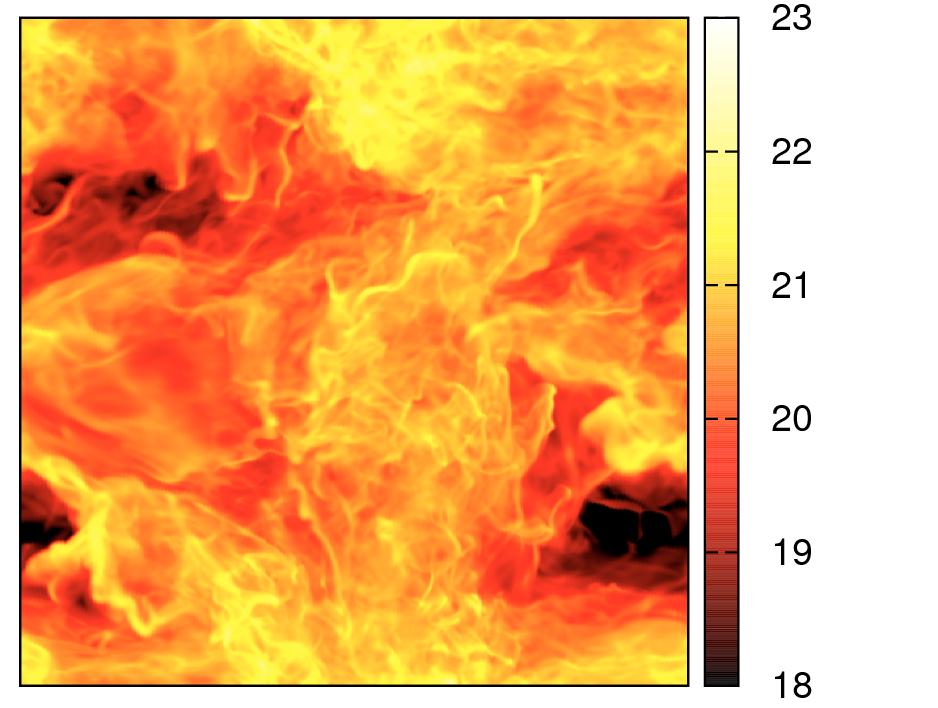}
\includegraphics[height=0.24\linewidth]{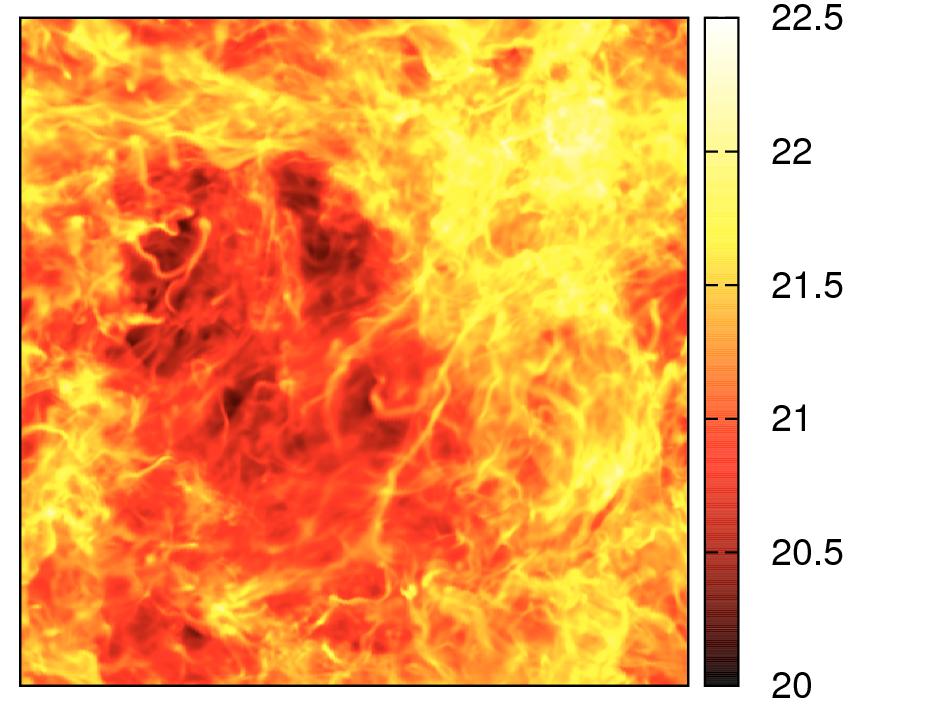}
\includegraphics[height=0.24\linewidth]{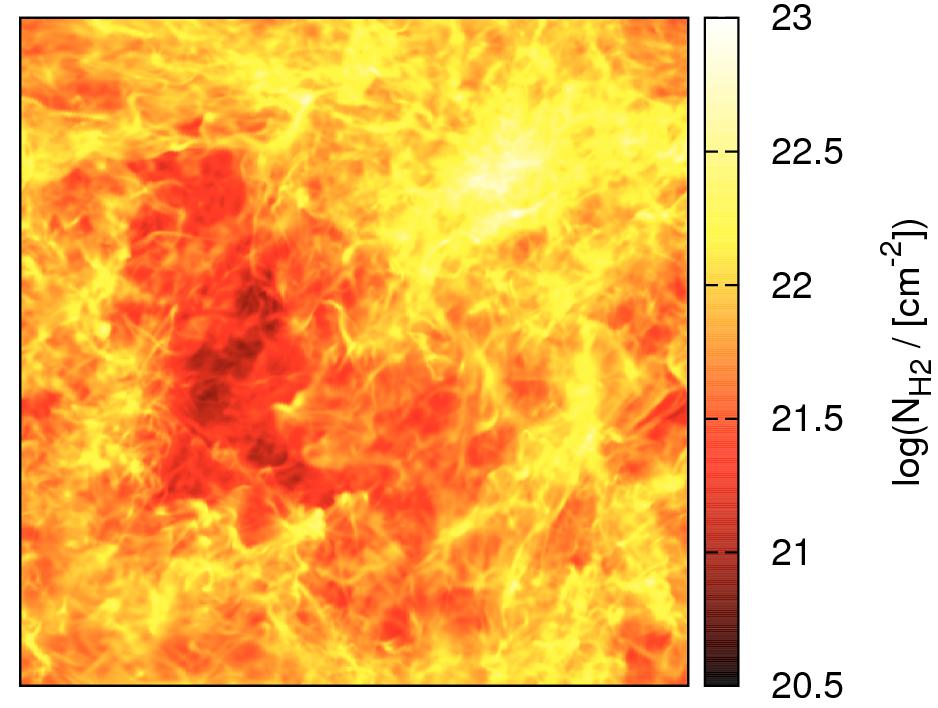}
}
\centerline{
\includegraphics[height=0.24\linewidth]{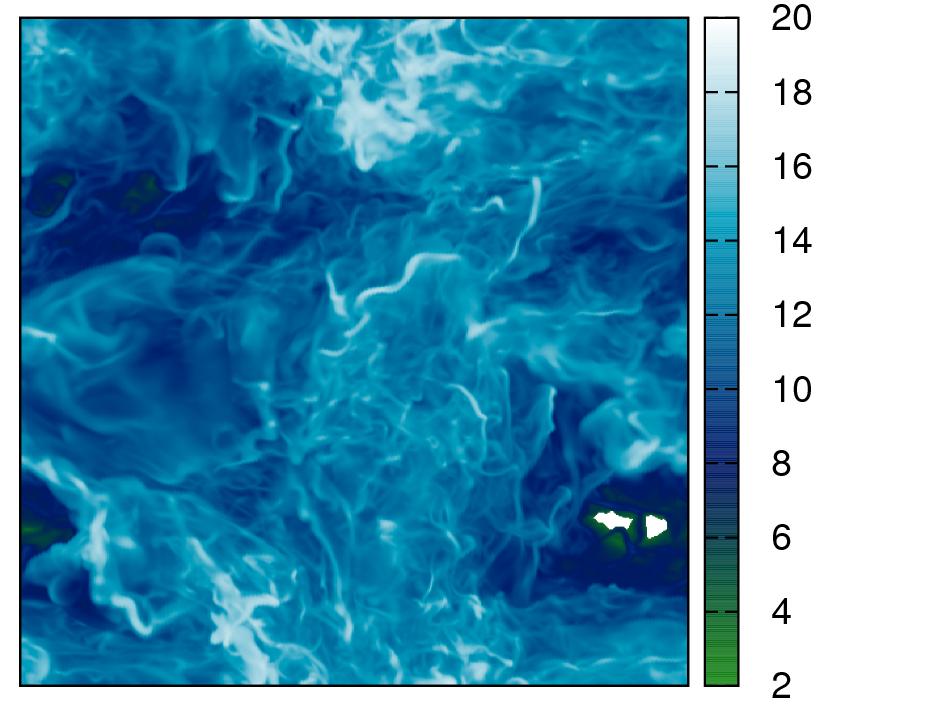}
\includegraphics[height=0.24\linewidth]{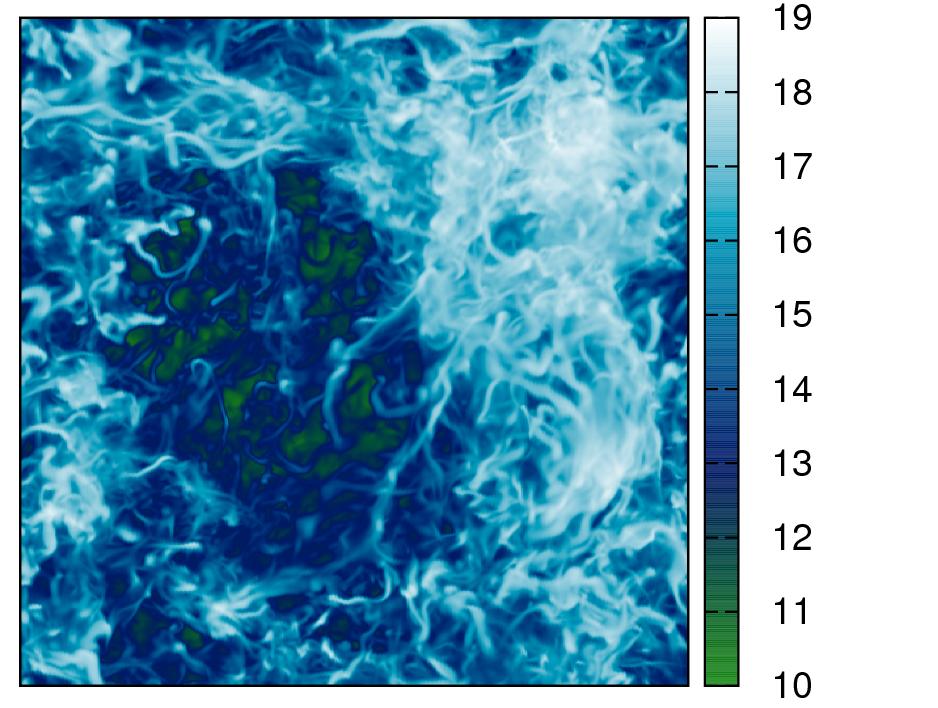}
\includegraphics[height=0.24\linewidth]{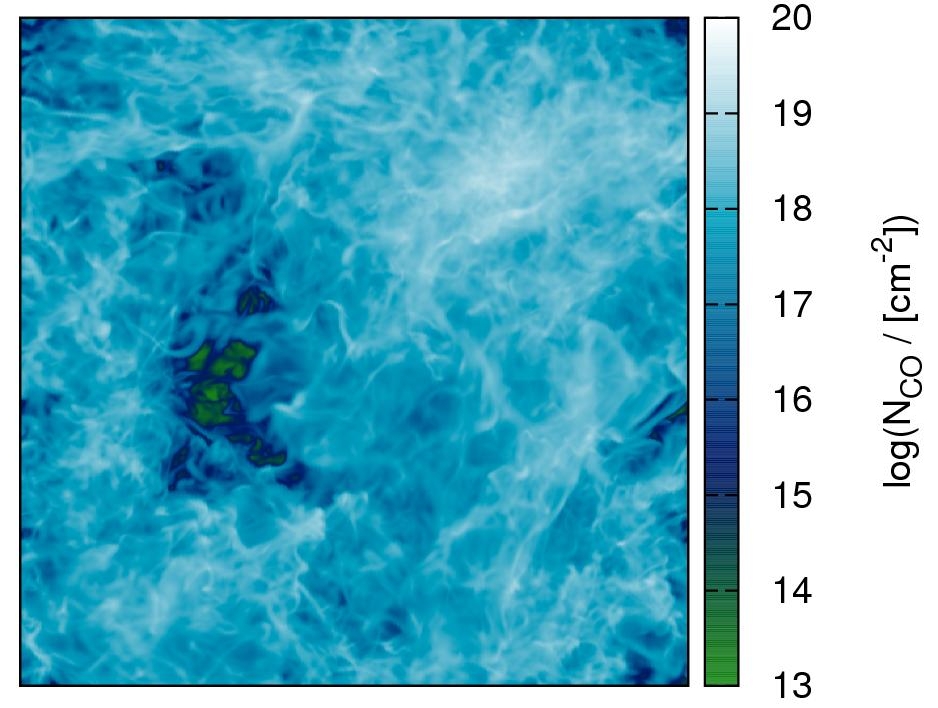}
}
\centerline{
\includegraphics[height=0.24\linewidth]{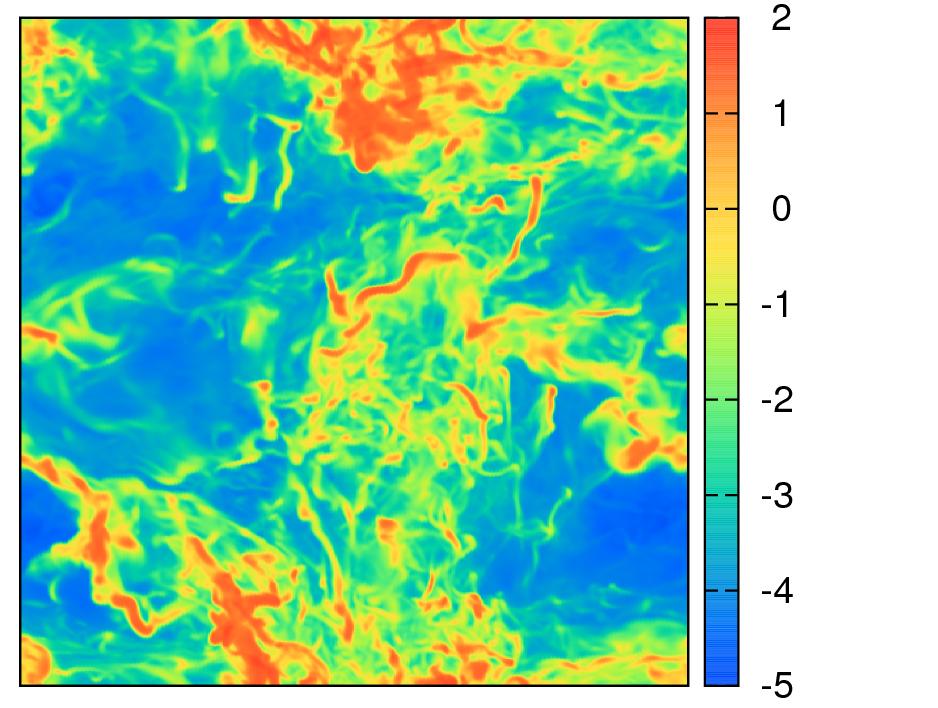}
\includegraphics[height=0.24\linewidth]{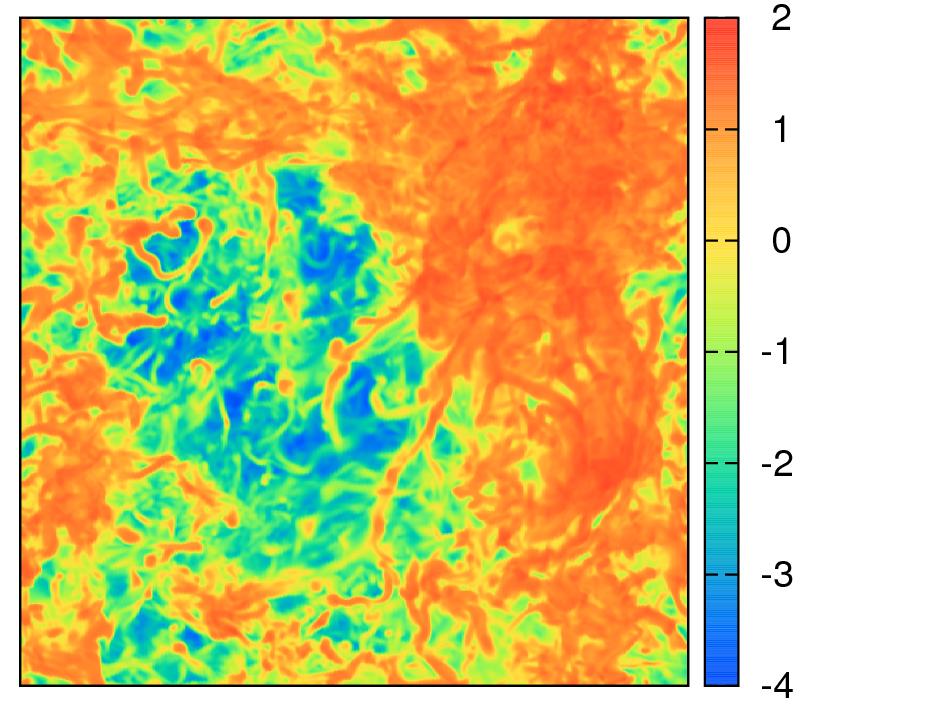}
\includegraphics[height=0.24\linewidth]{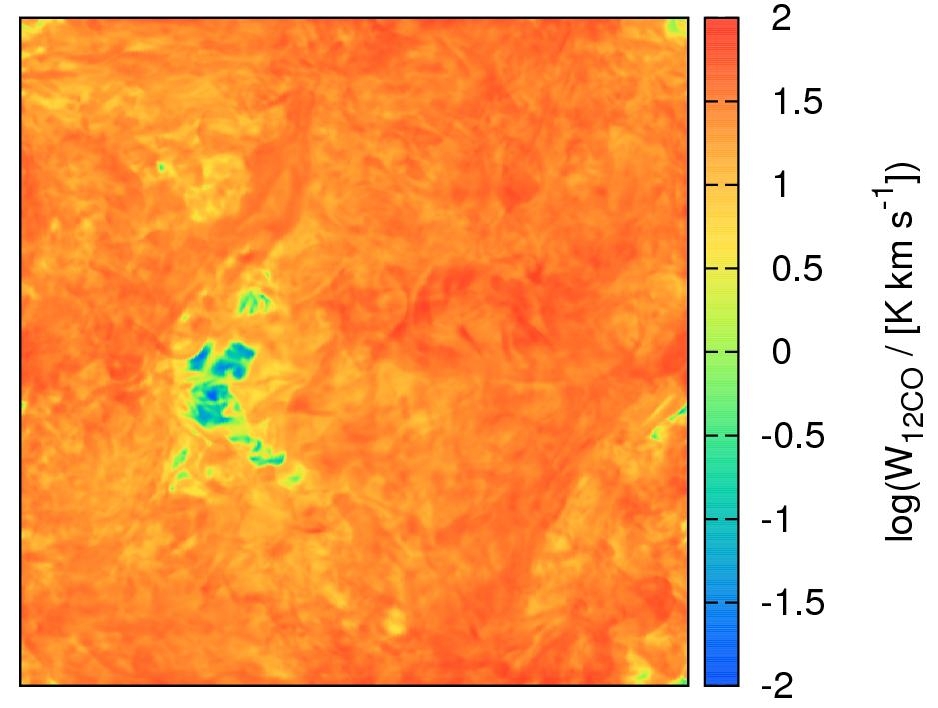}
}
\centerline{
\includegraphics[height=0.24\linewidth]{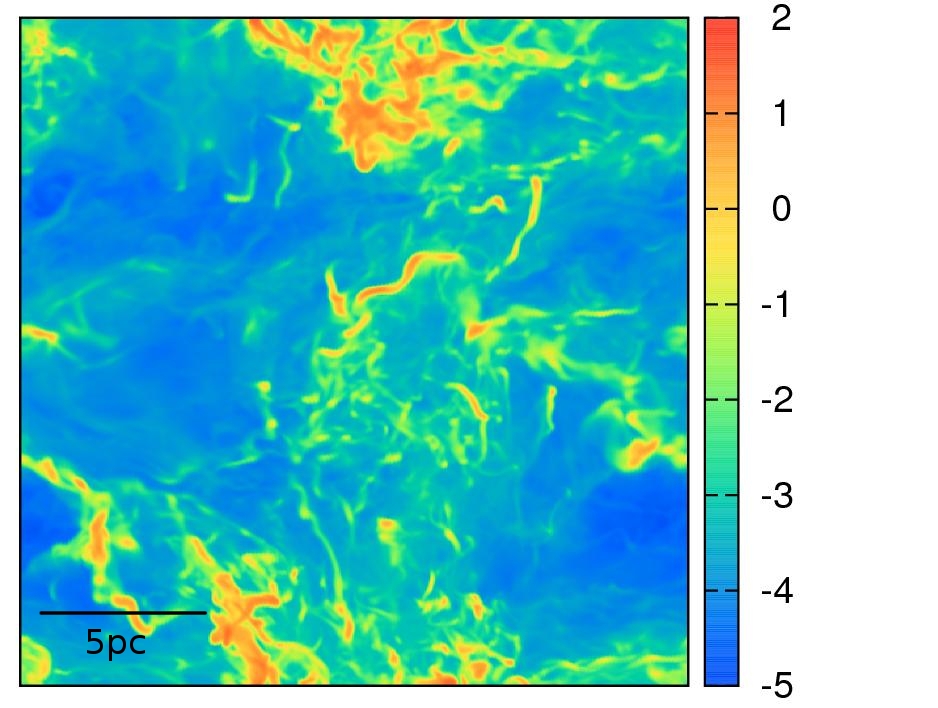}
\includegraphics[height=0.24\linewidth]{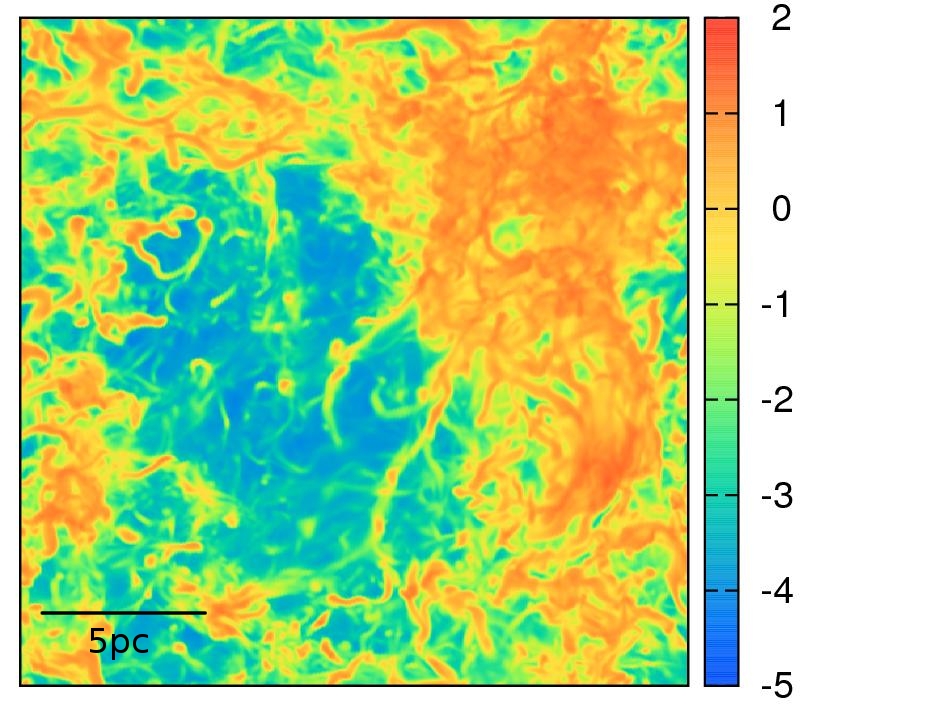}
\includegraphics[height=0.24\linewidth]{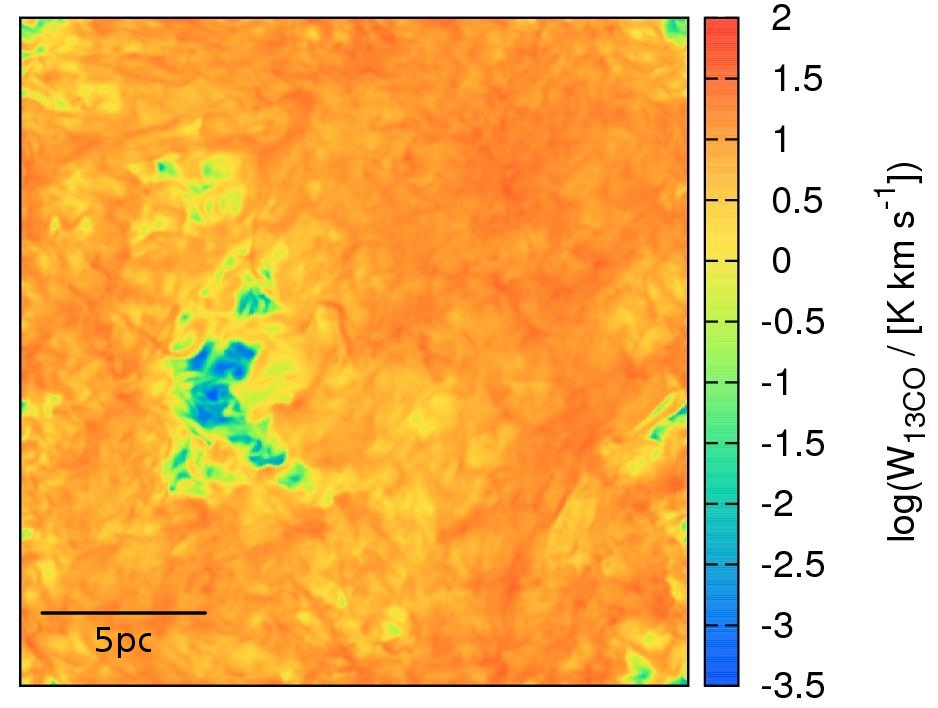}
}
\caption{Logarithmic maps of column density as well as velocity-integrated intensity along the $z$-direction for our n30 (left column), n100 (middle column) and n300 models (right column). From top to bottom we show the different chemical components: total, H$_2$ and CO column density as well as the integrated intensity of $^{12}$CO and $^{13}$CO in the $J=1 \rightarrow 0$ transition. Each side of the simulation domain has a length of 20\,pc. Note that the velocity field of the n30 model uses a different turbulent random seed than the n100 and the n300 model. Furthermore, we caution the reader that our color bars use a different scaling in the individual plots.}
\label{fig:CDimages}
\end{figure*}

\begin{figure*}
\centerline{
\includegraphics[height=0.26\linewidth]{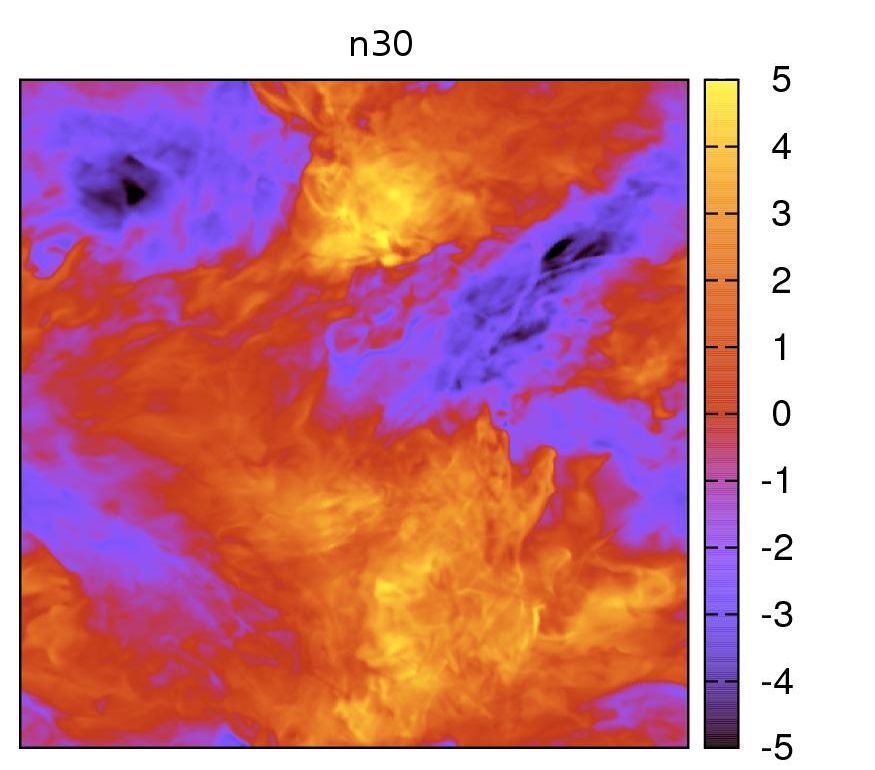}
\includegraphics[height=0.26\linewidth]{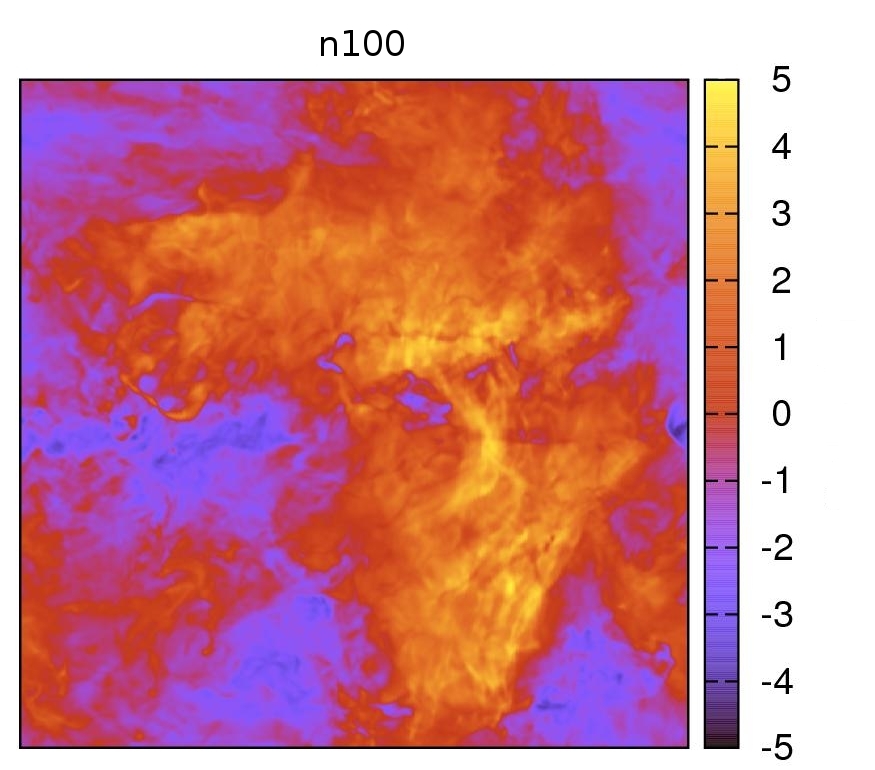}
\includegraphics[height=0.26\linewidth]{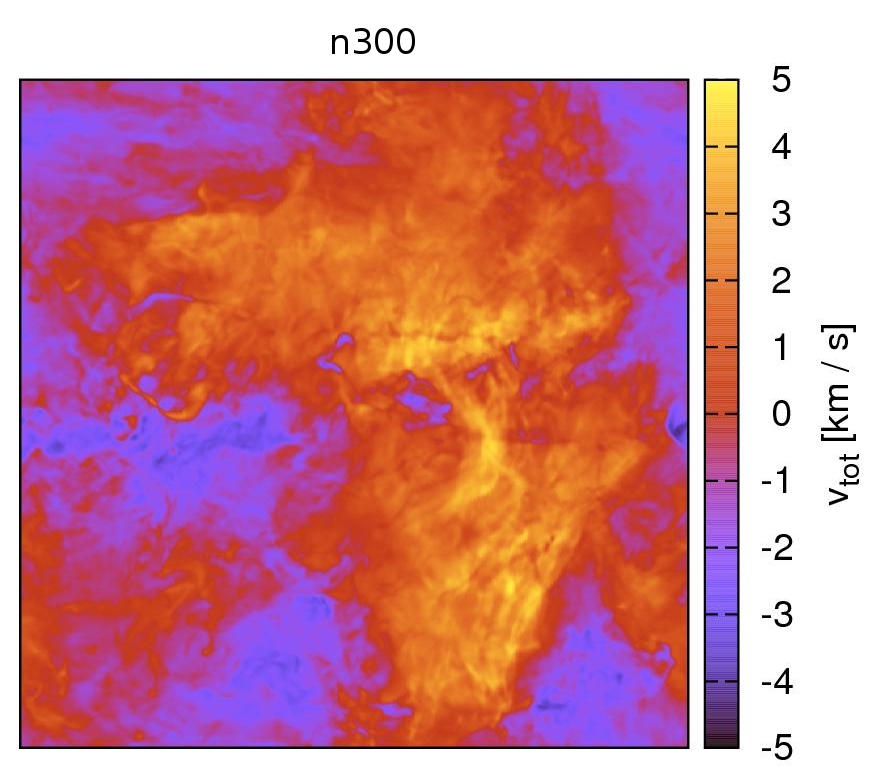}
}
\centerline{
\includegraphics[height=0.24\linewidth]{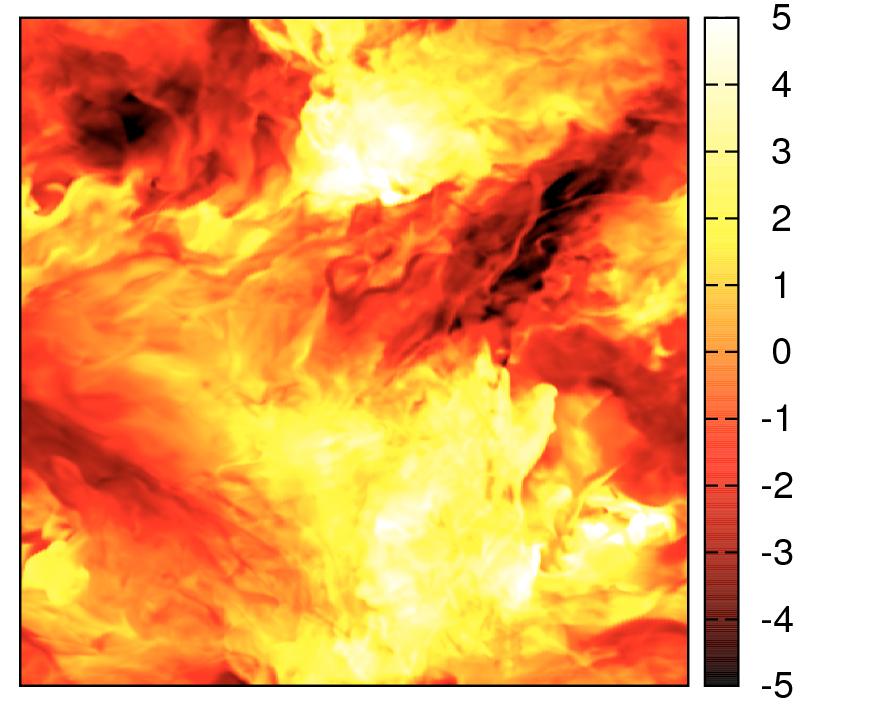}
\includegraphics[height=0.24\linewidth]{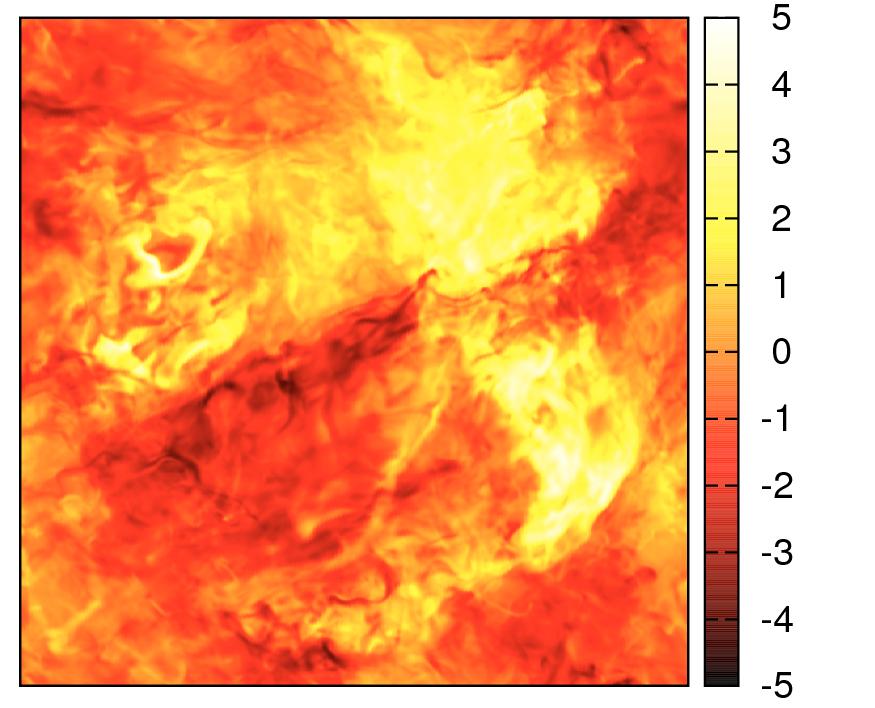}
\includegraphics[height=0.24\linewidth]{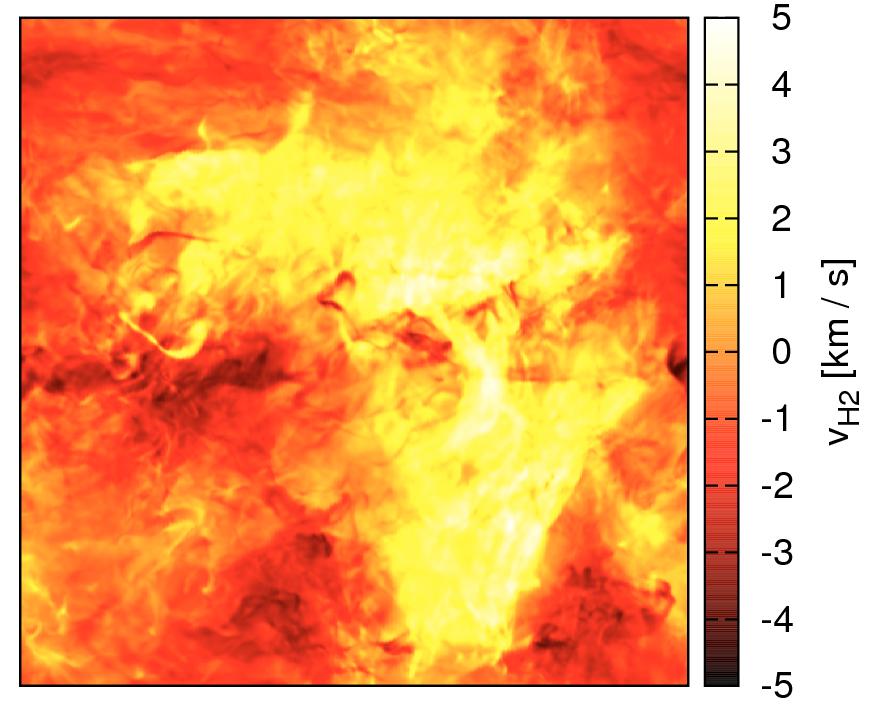}
}
\centerline{
\includegraphics[height=0.24\linewidth]{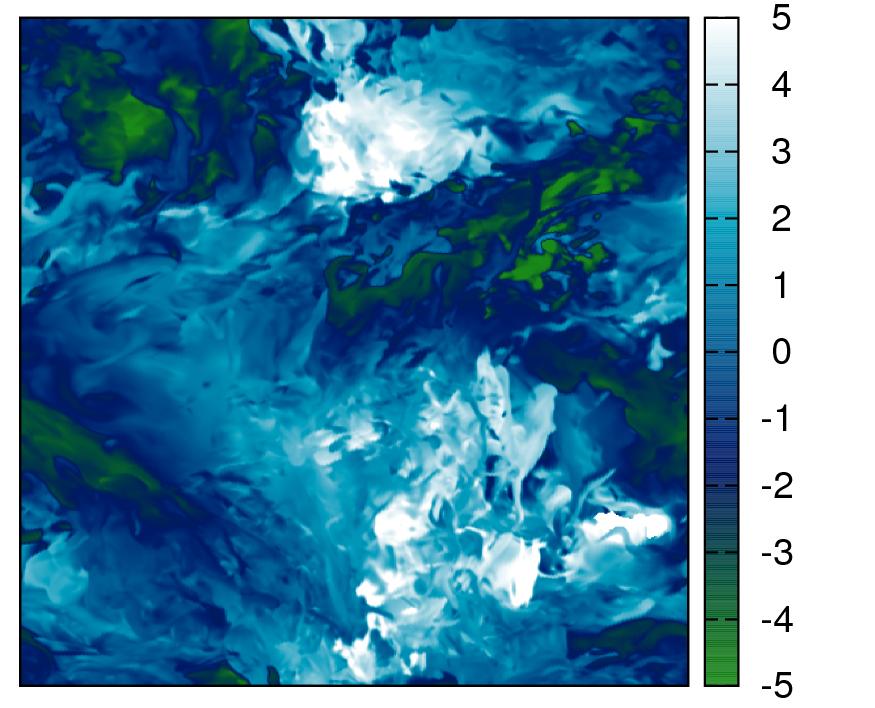}
\includegraphics[height=0.24\linewidth]{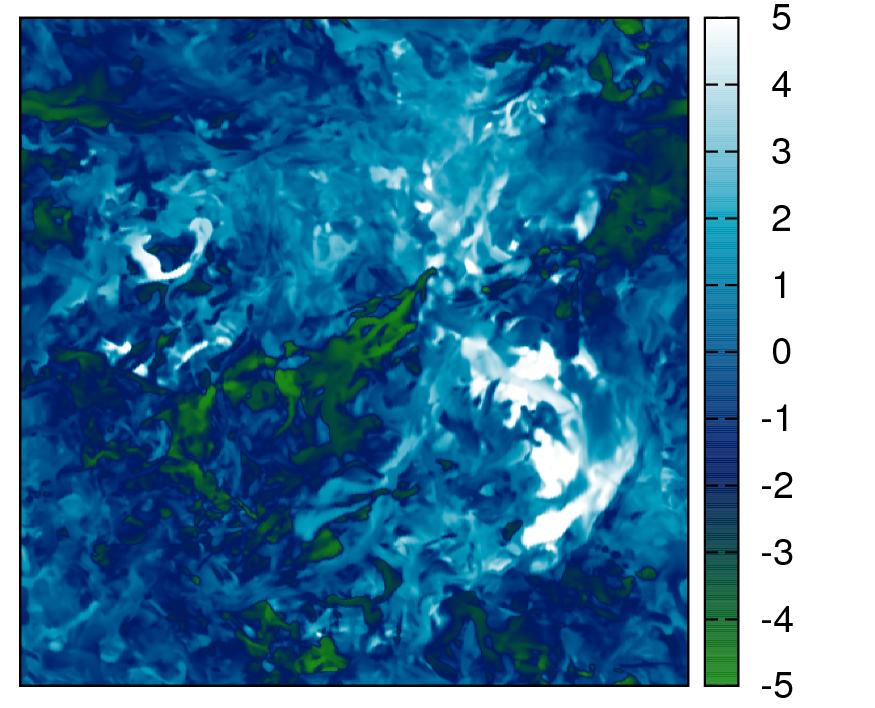}
\includegraphics[height=0.24\linewidth]{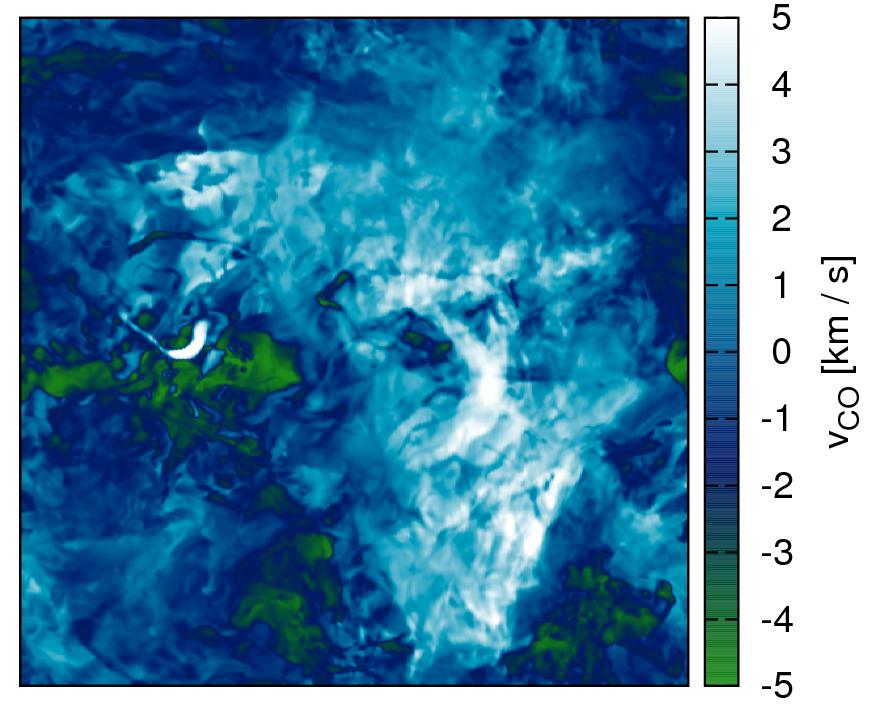}
}
\centerline{
\includegraphics[height=0.24\linewidth]{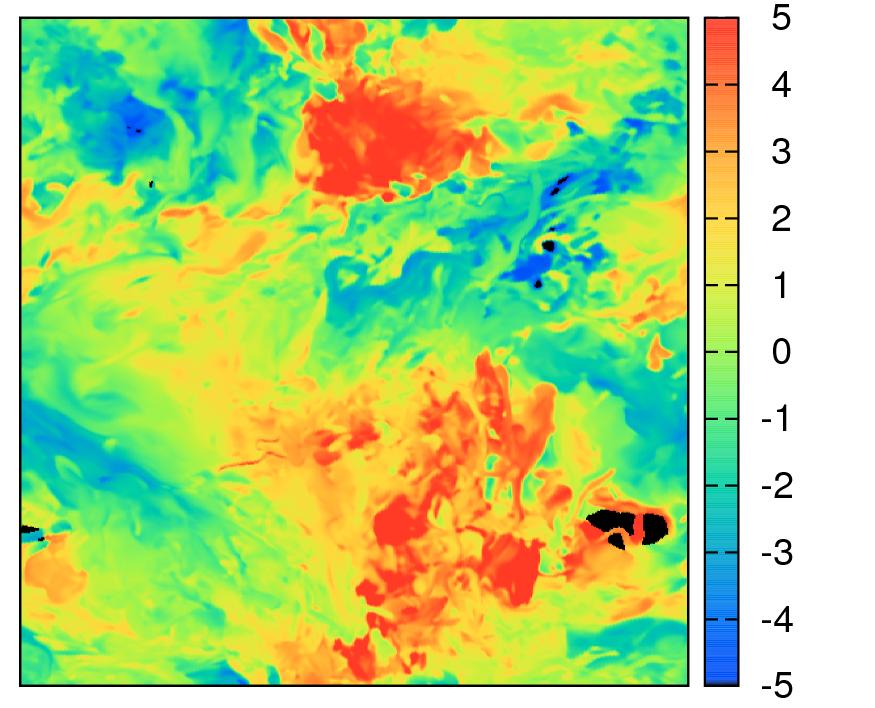}
\includegraphics[height=0.24\linewidth]{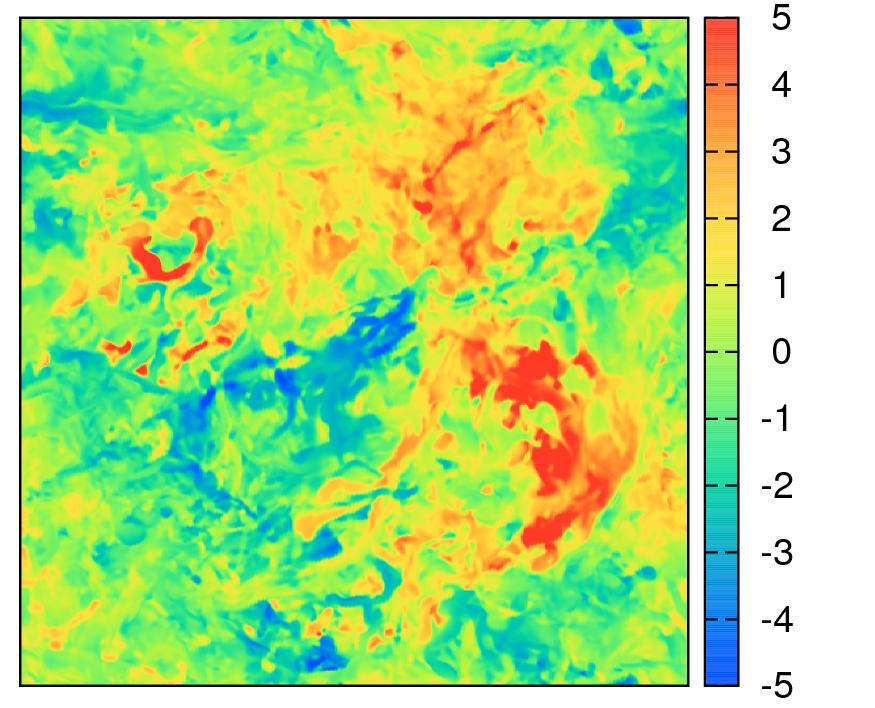}
\includegraphics[height=0.24\linewidth]{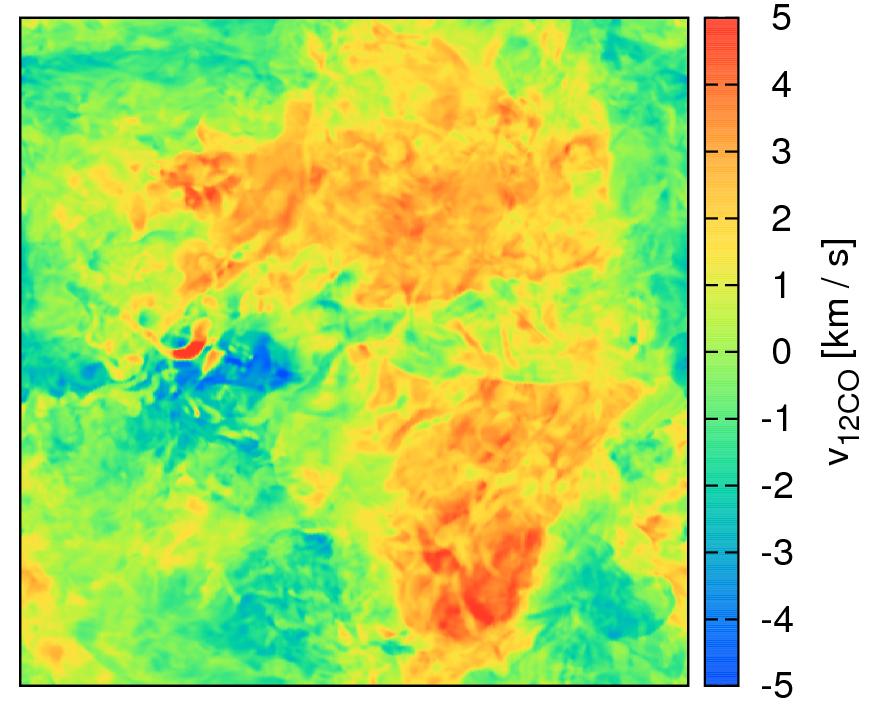}
}
\centerline{
\includegraphics[height=0.24\linewidth]{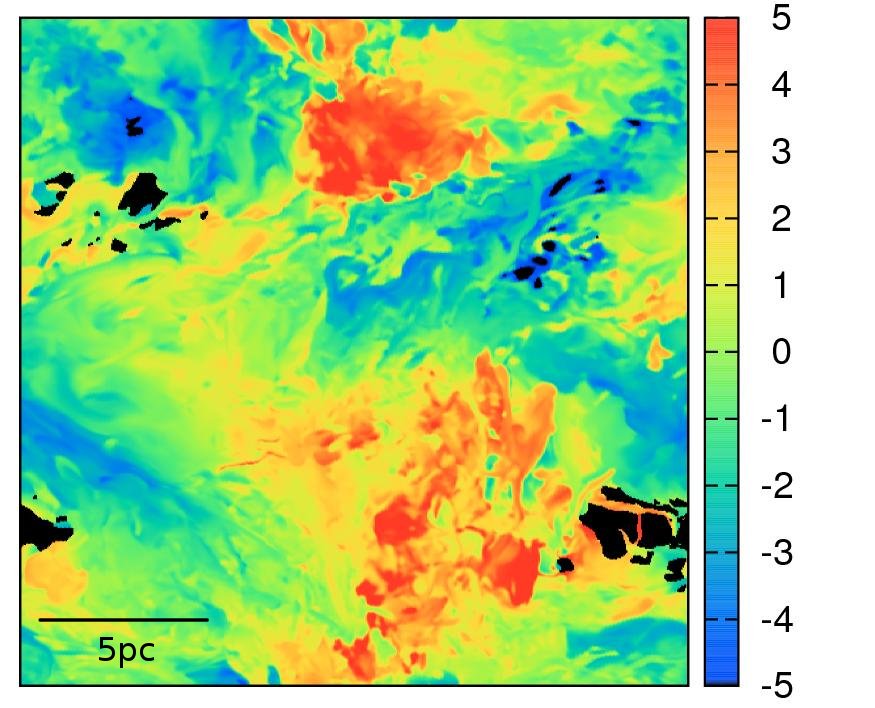}
\includegraphics[height=0.24\linewidth]{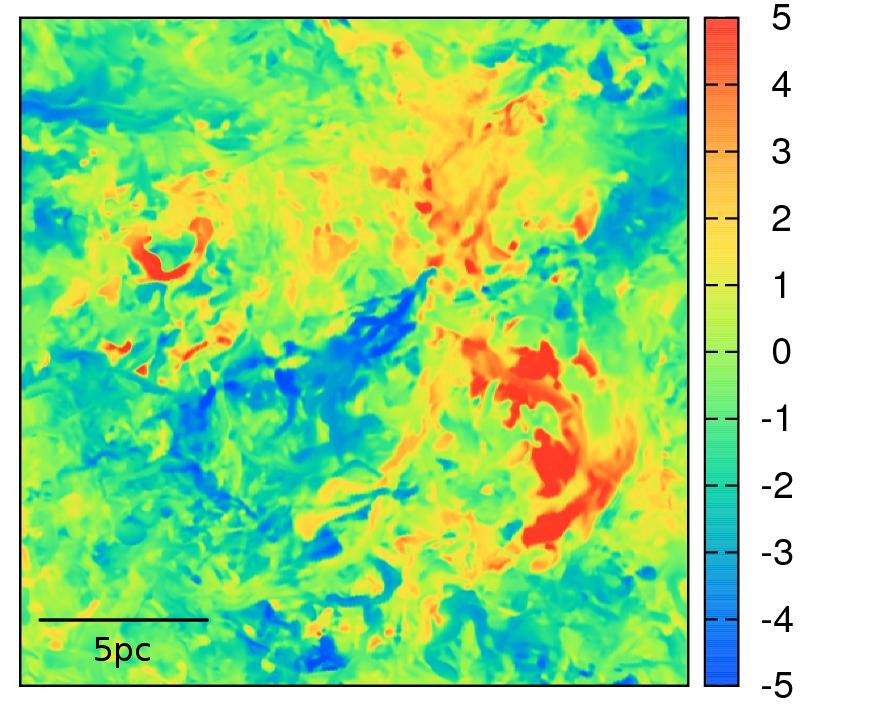}
\includegraphics[height=0.24\linewidth]{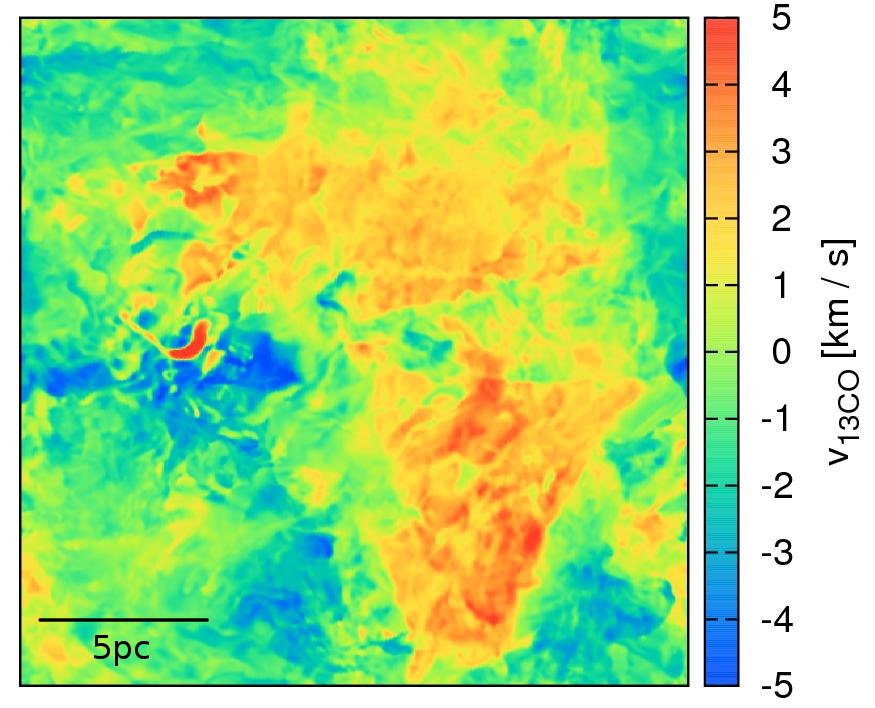}
}
\caption{Same as Fig. \ref{fig:CDimages}, but with maps of centroid velocities (CV). Black areas in the $^{12}$CO and $^{13}$CO intensity map of the n30 model denote regions where the brightness temperatures are zero along the line-of-sight. We mask these regions, because no proper centroid velocities can be computed there.}
\label{fig:images}
\end{figure*}

We perform numerical simulations and apply radiative transfer post-processing to our data in order to analyze the influence of chemical inhomogeneities and optical depth effects on the $\Delta$-variance analysis. Table~\ref{tab:setup} gives an overview of our numerical models. Mass- and volume-weighted quantities are defined via $\langle f \rangle_{\text{mass}} = \sum f \rho \text{d}V / \sum \rho \text{d}V$ and $\langle f \rangle_{\text{vol}} = \sum f \text{d}V / \sum \text{d}V$, respectively. For more information about the H$_2$ and CO distributions produced in this kind of turbulent simulation as well as integrated intensity and column density PDFs, we refer the reader to \citet{GloverEtAl2010} and \citeauthor{ShettyEtAl2011a}~(2011a).

Fig. \ref{fig:CDimages} and \ref{fig:images} show the column density and integrated intensity maps as well as the centroid velocity maps computed via equations (\ref{eq:CV}), (\ref{eq:W}) and (\ref{eq:N}) for all models and chemical components. Regarding Fig. \ref{fig:CDimages}, we find that the total and H$_2$ density models show similar structures on all spatial scales, which is because most of the hydrogen is in molecular form at this time (see also Table~\ref{tab:setup}). The CO column densities also trace similar structures, but span a much wider range of values, demonstrating that carbon monoxide has very low abundances along the low column density LoS. Furthermore, the $^{12}$CO and $^{13}$CO intensity maps also largely reflect the distribution of the carbon monoxide gas. We find that the intensity maps are much smoother and span a smaller range of values ($4-7$ orders of magnitude in integrated intensity compared to $7-18$ orders of magnitude in column density), which is due to the fact that the $J=1 \rightarrow 0$ line of $^{12}$CO is easily excited and can be bight even in low-density cloud regions. In the case of the $^{13}$CO intensity maps, the abundance is lower by a factor of $R_{12/13} = 50$ compared to the $^{12}$CO intensity models. Nevertheless, $^{13}$CO is optically thin and so the peaks in the $^{13}$CO maps coincide with those of the CO density models.

\subsection{$\Delta$-variance analysis of the CV maps}
\label{subsec:slopesDV}

\begin{figure}
\centerline{
\includegraphics[width=0.62\linewidth]{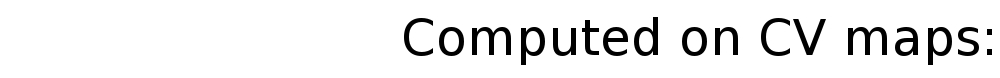} \\
}
\centerline{
\includegraphics[width=0.10\linewidth]{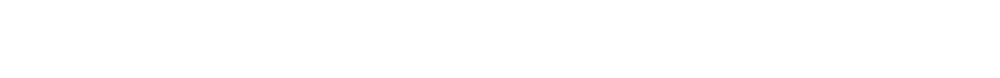} \\
}
\centerline{
\includegraphics[height=0.73\linewidth,width=1.0\linewidth]{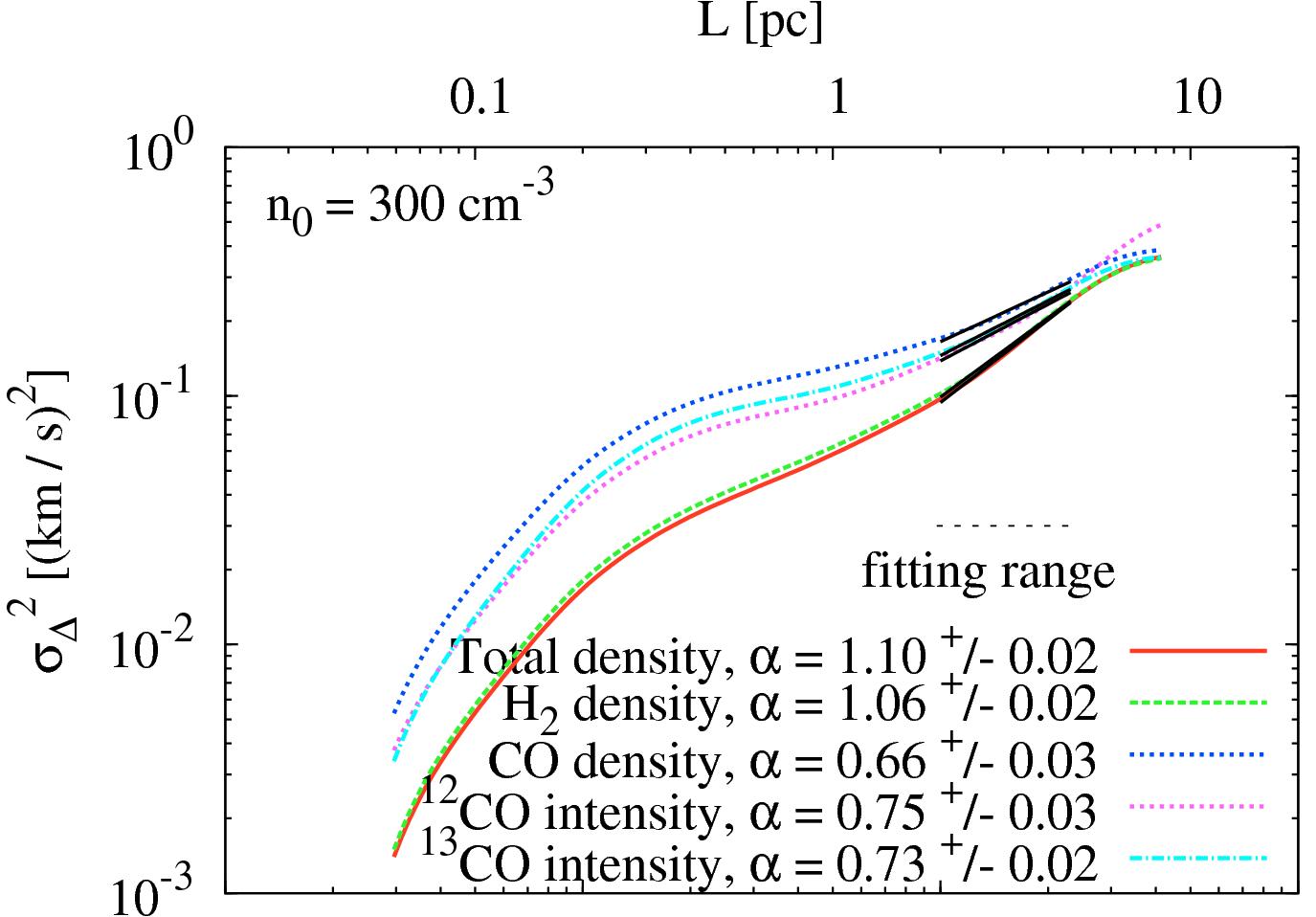} \\
}
\centerline{
\includegraphics[height=0.65\linewidth,width=1.0\linewidth]{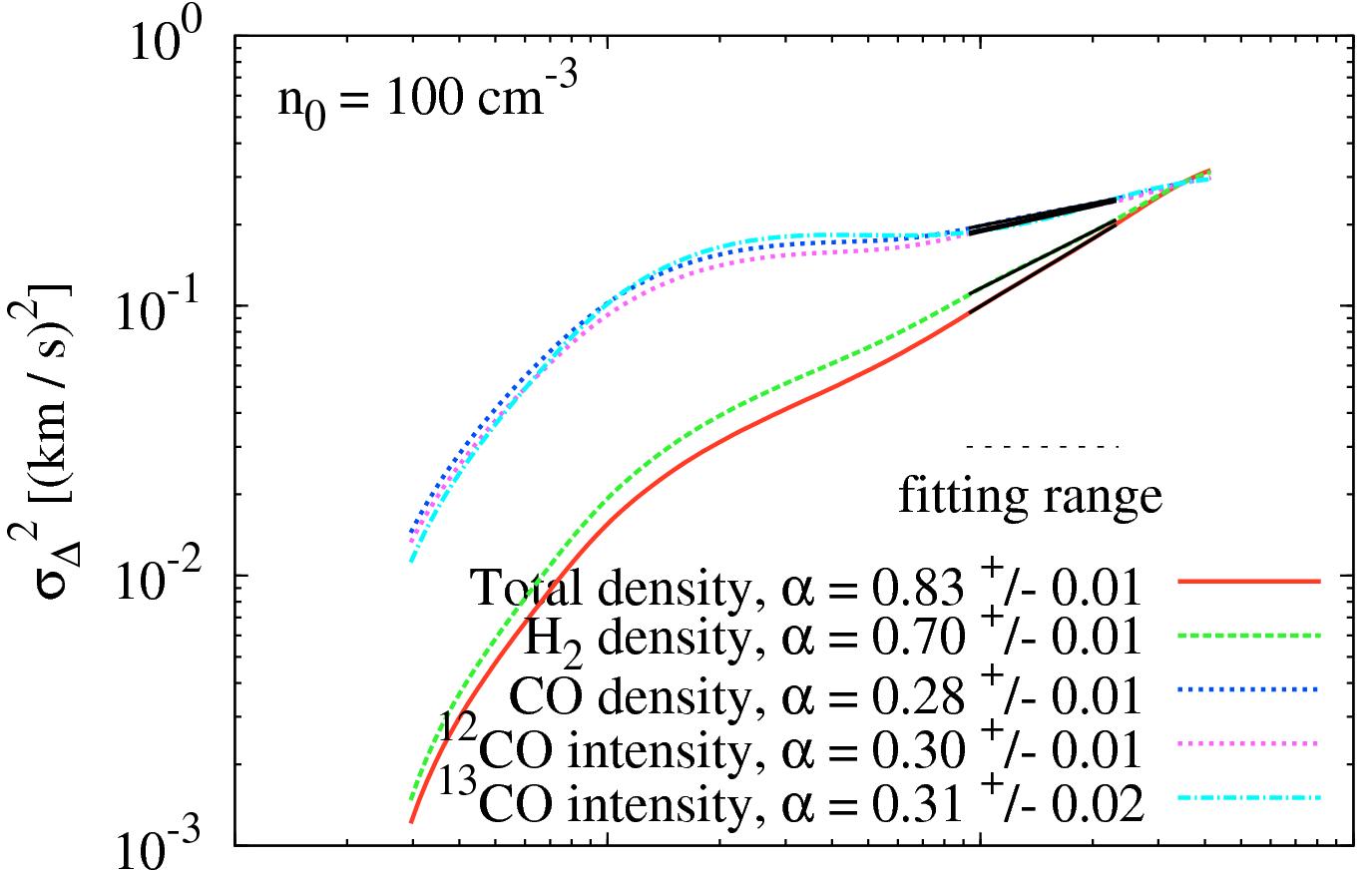} \\
}
\centerline{
\includegraphics[height=0.75\linewidth,width=1.0\linewidth]{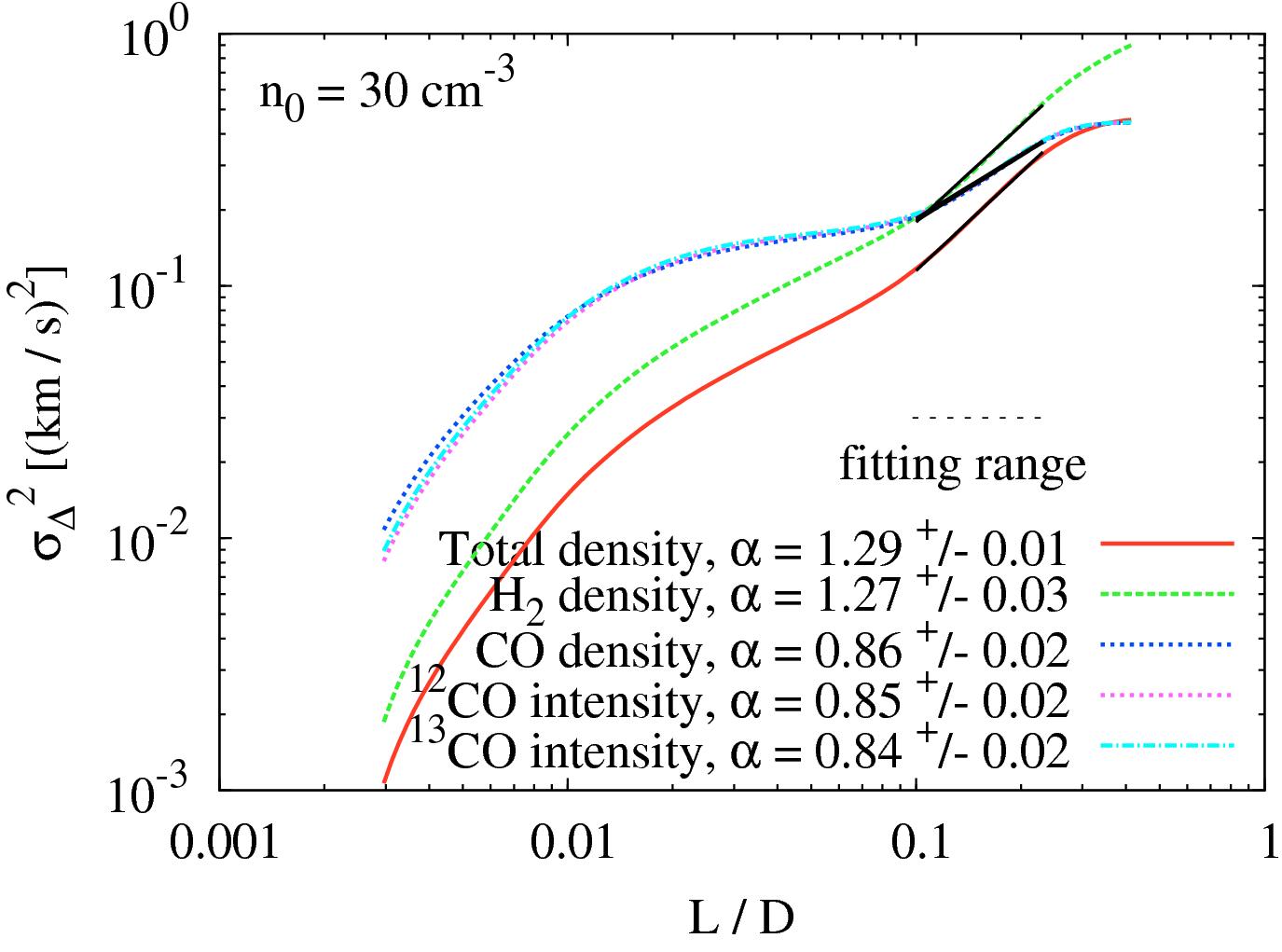} \\
}
\caption{$\Delta$-variance spectra as a function of the spatial scale, averaged over all available time snapshots and the different lines-of-sights $x$, $y$ and $z$. The spatial scale is normalized by the total box size. From top to bottom: spectra for our three different density models, i.e. n300, n100 and n30, computed for the centroid velocity (CV) maps. In each plot we show the $\Delta$-variance spectra for our various chemical components, i.e. for the total density, H$_2$ and CO density as well as for the $^{12}$CO and $^{13}$CO intensity. We use a fitting range from $1/10$ to $1/4$ of the total box size ($0.1 \lesssim \ell/D \lesssim 0.25$), as constrained by \citet{FederrathEtAl2010} and \citet{KonstandinEtAl2012a}, which is indicated by a horizontal dashed line and is the same in each density model. The different power-law functions in the fitting range are indicated with a black solid line on each spectrum. Furthermore, we list the slopes $\alpha$ and their errors from a $\chi^2$-fit in each plot.}
\label{fig:DVspectra}
\end{figure}

\begin{table*}
\begin{tabular}{l|c|c|c|c|c}
\hline\hline
 & Total density & H$_2$ density & CO density & $^{12}$CO intensity & $^{13}$CO intensity \\
 & $\alpha$ & $\alpha$ & $\alpha$ & $\alpha$ & $\alpha$ \\
\hline
n300 & $1.10 \pm 0.02$ & $1.06 \pm 0.02$ & $0.66 \pm 0.03$ & $0.75 \pm 0.03$ & $0.73 \pm 0.02$ \\
n100 & $0.83 \pm 0.01$ & $0.70 \pm 0.01$ & $0.28 \pm 0.01$ & $0.30 \pm 0.01$ & $0.31 \pm 0.02$ \\
n30 & $1.29 \pm 0.01$ & $1.27 \pm 0.03$ & $0.86 \pm 0.02$ & $0.85 \pm 0.02$ & $0.84 \pm 0.02$ \\
\hline
\end{tabular}
\caption{Spectral slope of the $\Delta$-variance spectrum, computed using centroid velocities weighted by the indicated quantity (see also Fig. \ref{fig:DVspectra}). The errors are computed by the $\chi^2$ fitting method. The slope $\alpha$ is related to the $\Delta$-variance via $\sigma_\Delta^2 (\ell) \propto \ell^{\alpha}$ in a linear regime. The values $\alpha$ can also be used to compute spectral slopes $\gamma = \alpha / 2$ for a linewidth-size relation $\sigma_\Delta (\ell) \propto \ell^{\gamma}$, readily comparable to observational measurements.}
\label{tab:DVslopes}
\end{table*}

We compute the $\Delta$-variance for all models (n300, n100 and n30) and all chemical tracers: the total density, H$_2$ density, CO density, $^{12}$CO and $^{13}$CO intensity. We average all spectra from 3 snapshots in time (with 3 line-of-sight directions each) where we can assume both the chemistry and the turbulence to be in a stationary and converged state. Fig. \ref{fig:DVspectra} shows $\Delta$-variance spectra as a function of spatial scale for all density models and chemical tracers, computed for the centroid velocity (CV) maps, as introduced in Section \ref{subsec:CV}. Table \ref{tab:DVslopes} summarizes all slope values $\alpha$ obtained from a $\chi^2$-fit for the $\Delta$-variance of the velocity field, $\sigma_\Delta^2 (\ell) \propto \ell^{\alpha}$, which are also listed in each plot in Fig. $\ref{fig:DVspectra}$.

We generally find significant differences between the various models and tracers. In our n300 model, we obtain similar slopes for the total density and the H$_2$ density (see Table \ref{tab:DVslopes}). Regarding the spectra of those two cases, we find an excellent agreement within the values of $\sigma_\Delta^2$ for both models, as shown in the top plot of Fig. \ref{fig:DVspectra}. For the low-density model n30 (bottom plot of Fig. \ref{fig:DVspectra}), we obtain a large discrepancy between the $\sigma_\Delta^2$ values of total density and the H$_2$ density model. This is because in this simulation the fraction of molecular gas is much smaller than in the higher density runs and so H$_2$ no longer follows the total gas density \citep{GloverEtAl2010}. The H$_2$ density is therefore more inhomogeneous than the gas density, and as a result the centroid velocities weighted by H$_2$ fluctuate more on all scales than those weighted by the total density. The n100 model makes up an intermediate case between the n300 and the n30 density models. In this model, the correlation between the total and the H$_2$ density is worse than in the n300 case, but still better than in the n30 model.

Comparing the total and H$_2$ density models to the various CO tracers, we find that the slopes of the former are significantly steeper compared to the slopes of the latter, independent of the density (see Table \ref{tab:DVslopes}). However, the $\Delta$-variance spectra of the various CO tracers show a good agreement with each other over nearly all spatial scales. Furthermore, we always find the $\sigma_\Delta^2$ values of the CO tracers to be significantly larger than those of the total and H$_2$ density. This is because CO is mainly located in denser regions of the cloud (see, e.g. \citeauthor{ShettyEtAl2011b}, 2011a,b), leading to higher density contrasts compared to the total and H$_2$ density and thus to larger variances in velocity space.

\subsection{$\Delta$-variance analysis of the intensity and column density maps}
\label{subsec:DVWN}

\begin{figure}
\centerline{
\includegraphics[width=0.95\linewidth]{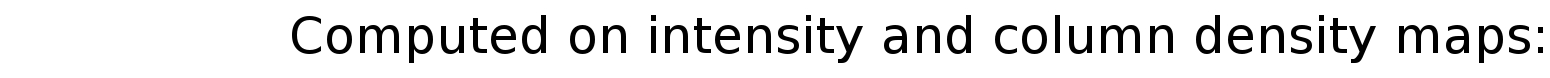} \\
}
\centerline{
\includegraphics[width=0.10\linewidth]{images/blank.jpg} \\
}
\centerline{
\includegraphics[height=0.73\linewidth,width=1.0\linewidth]{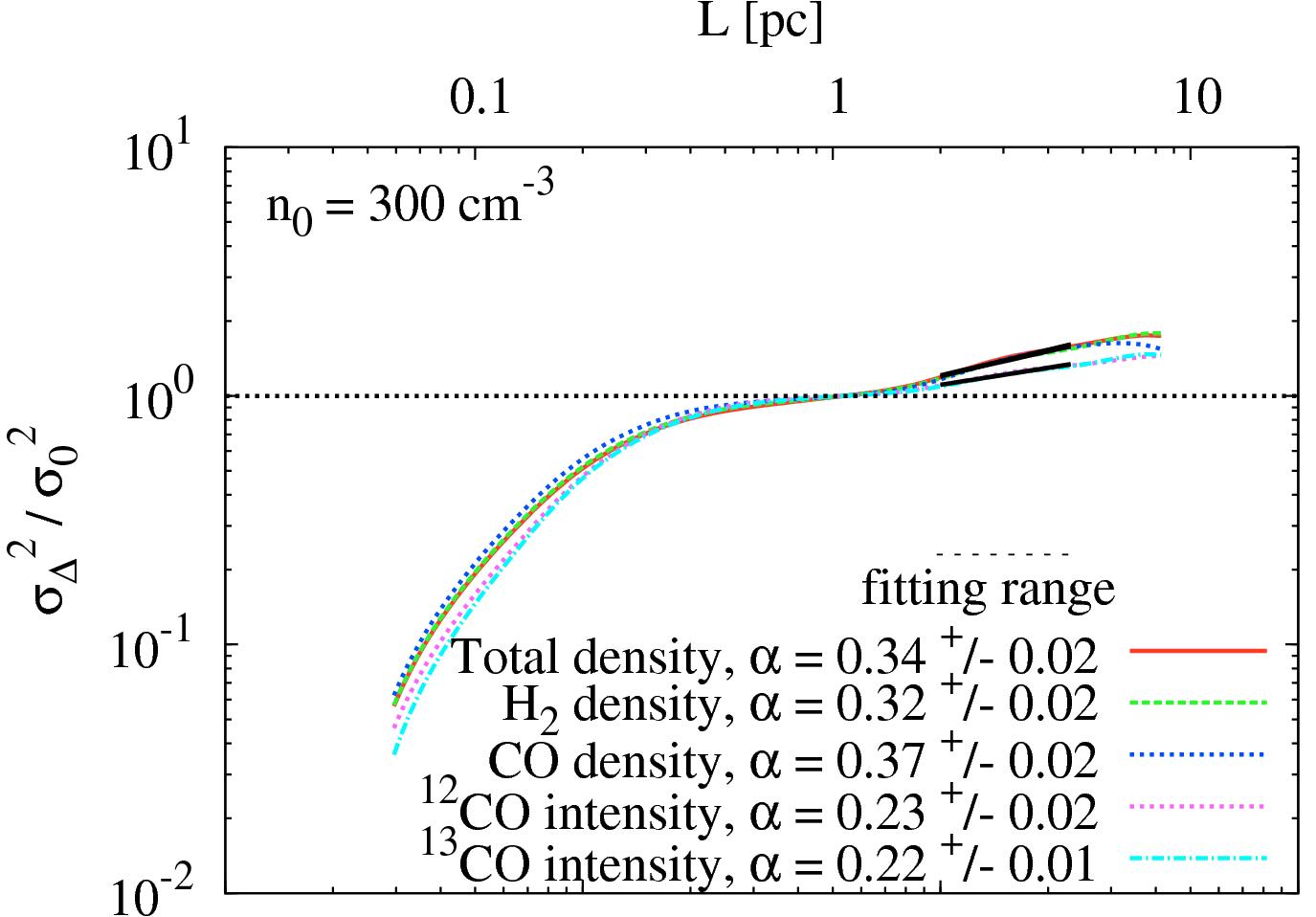} \\
}
\centerline{
\includegraphics[height=0.65\linewidth,width=1.0\linewidth]{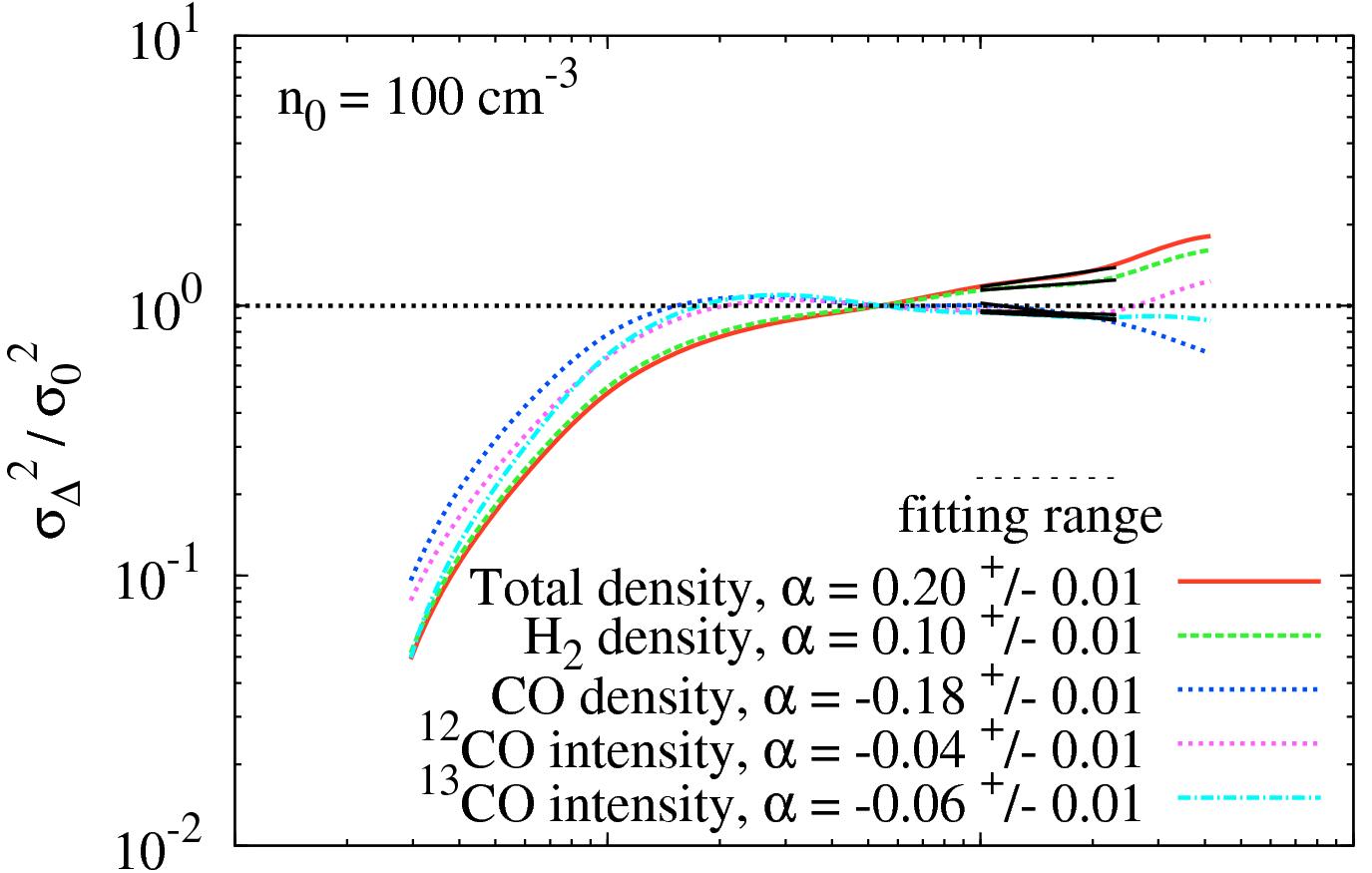} \\
}
\centerline{
\includegraphics[height=0.75\linewidth,width=1.0\linewidth]{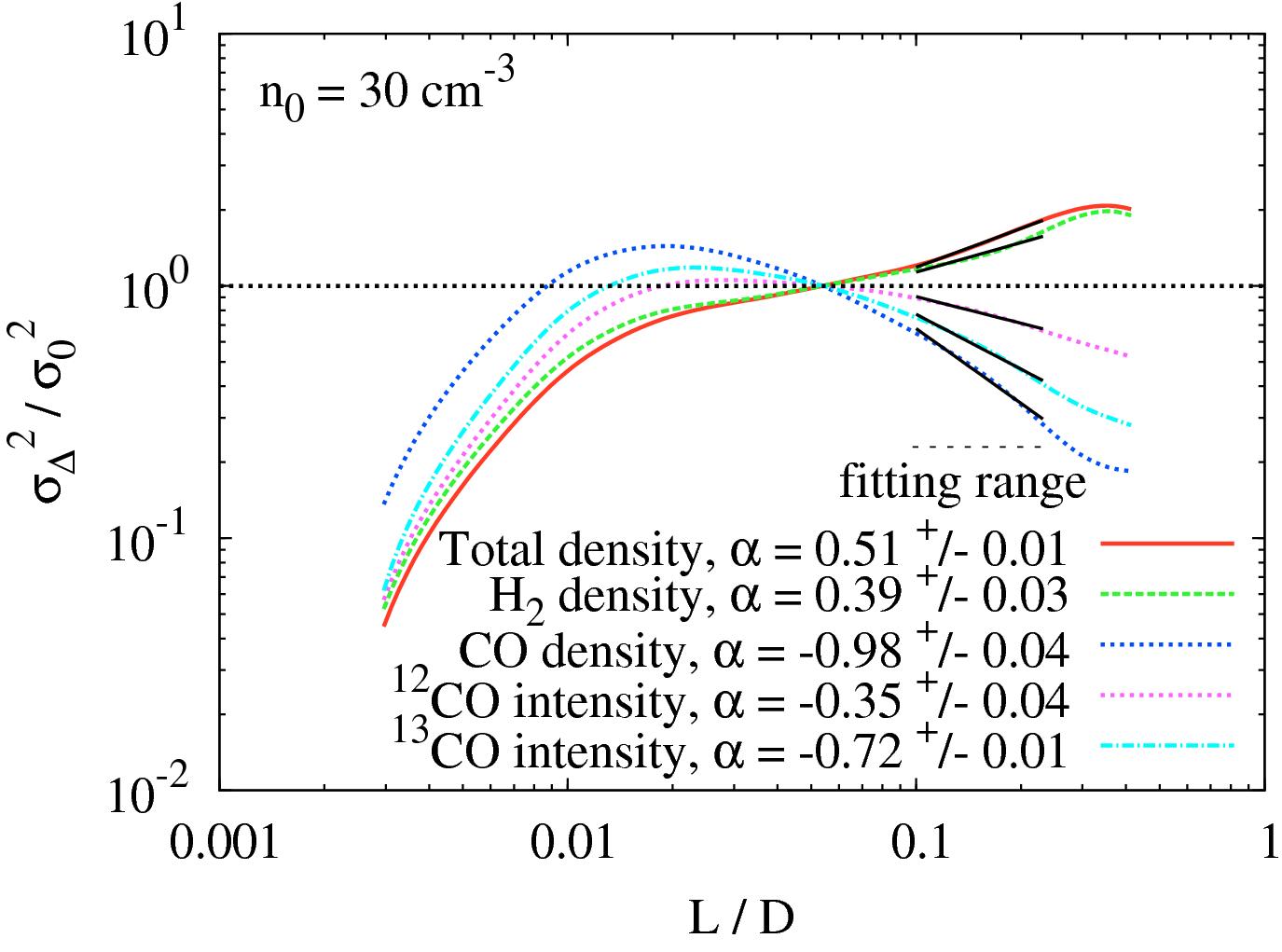} \\
}
\caption{Same as Fig. \ref{fig:DVspectra}, but computed using maps of integrated intensity and column density. In this plot, we normalize each curve by the corresponding value $\sigma_0^2$ measured at an arbitrary spatial scale of 5\% of the total box, in order to better compare the spectra computed for the intensity and column density maps with each other. For a better visualization, we also show a horizontal dashed line at a value of $\sigma_\Delta^2 / \sigma_0^2 = 1$ in each plot. Furthermore, for clarity, we do not plot the error bars of the individual spectra here. The slopes $\alpha$ and their errors from a $\chi^2$-fit are listed in each plot for all our chemical components.}
\label{fig:DVWNspectra}
\end{figure}

\begin{table*}
\begin{tabular}{l|c|c|c|c|c}
\hline\hline
 & Total density & H$_2$ density & CO density & $^{12}$CO intensity & $^{13}$CO intensity \\
 & $\alpha$ & $\alpha$ & $\alpha$ & $\alpha$ & $\alpha$ \\
\hline
n300 & $0.34 \pm 0.02$ & $0.32 \pm 0.02$ & $0.37 \pm 0.02$ & $0.23 \pm 0.02$ & $0.22 \pm 0.01$ \\
n100 & $0.20 \pm 0.01$ & $0.10 \pm 0.01$ & $-0.18 \pm 0.01$ & $-0.04 \pm 0.01$ & $-0.06 \pm 0.01$ \\
n30 & $0.51 \pm 0.01$ & $0.39 \pm 0.03$ & $-0.98 \pm 0.04$ & $-0.35 \pm 0.04$ & $-0.72 \pm 0.01$ \\
\hline
\end{tabular}
\caption{Same as Table \ref{tab:DVslopes}, but for the $\Delta$-variance computed directly from maps of integrated intensity and column density, as defined in equation (\ref{eq:W}) and (\ref{eq:N}). The corresponding $\Delta$-variance spectra are shown in Fig. \ref{fig:DVWNspectra}.}
\label{tab:DVWNslopes}
\end{table*}

Beyond the $\Delta$-variance analysis of centroid velocities, we can also apply the same method to any other quantities defined on the $x$-$y$ plane. In particular, it is possible to carry out a similar analysis for the total, H$_2$ and CO column densities and the $^{12}$CO and $^{13}$CO integrated intensities. The results of this analysis are illustrated in Fig. \ref{fig:DVWNspectra}. In contrast to Fig. \ref{fig:DVspectra}, we normalize each curve by the corresponding $\Delta$-variance value $\sigma_0^2$ measured at an arbitrary spatial scale of 5\% of the total box, in order to better compare the spectra computed for the intensity and column density maps with each other. Furthermore, we also show a horizontal dashed line at a value of $\sigma_\Delta^2 / \sigma_0^2 = 1$ in each plot. Table \ref{tab:DVWNslopes} summarizes all slope values $\alpha$ obtained from a $\chi^2$-fit for the individual $\Delta$-variance spectra.

As for the CV maps, we find significant differences between the spectra of the various models and tracers computed using the maps of integrated intensity and column density directly. In the high-density n300 model, we find a good correlation between the total density and the H$_2$ density (see also the slopes in Table \ref{tab:DVWNslopes}), for the same reasons as described in Section \ref{subsec:slopesDV}. Similarly, the low-density n30 model shows again the largest discrepancy. This is because the fraction of molecular gas is significantly smaller than in the n300 simulation, leading to different values of $\alpha$. The n100 model again defines an intermediate case between the n300 and the n30 density runs. In this model, the H$_2$ density better correlates with the total density than in the low-density n30 run, but still worse than in the high-density n300 run.

Regarding the simulations of the various CO tracers, we obtain very different scaling properties compared to the total and the H$_2$ density simulations, depending on the mean density. In the high-density n300 model, we find a good correlation between the two CO intensity cases, while the slopes of the total and CO density models are significantly steeper. However, both the total and H$_2$ density as well as the various CO tracer models show the same trend in $\sigma_\Delta^2$ over all spatial scales, having $\alpha > 0$. This situation changes in the low-density n30 model, and we find strong differences between the CO tracers and the total and H$_2$ density runs over all spatial scales. While the structure in the computational domain on larger scales for the total and H$_2$ column density can still be described by a slope $\alpha > 0$, the $\Delta$-variance spectra for the CO tracers have negative slopes in this density run (see Table \ref{tab:DVWNslopes}). A similar situation as in the low-density n30 simulation can be seen in the intermediate n100 density model. While we again measure $\alpha > 0$ for the total and H$_2$ density cases, we obtain $\alpha \lesssim 0$ for the CO tracers, meaning that the spatial distribution of observable carbon monoxide is completely different compared to the distribution of the H$_2$ gas in the n30 and the n100 density simulations.

Looking at Fig. \ref{fig:DVWNspectra}, we see that the reason that we obtain negative values for $\alpha$ for the CO tracers in the low-density runs is that the corresponding $\Delta$-variance spectra turn over at relatively low spatial scales. This indicates that the CO in these simulations has a characteristic spatial scale of approximately 2\% of the total box size, corresponding to around 0.4\,pc. This is consistent with what we see in the CO column density projections and integrated intensity maps shown in Fig. \ref{fig:CDimages}. In the low-density n30 simulation, CO is sufficiently well shielded to resist photodissociation only in the dense filaments formed by turbulent compressions. These structures are thin and contribute strongly to the $\Delta$-variance on scales comparable to their width. Moreover, the fact that this characteristic scale is roughly the same in all CO spectra demonstrates that the radiative transfer post-processing does not significantly change the characteristic spatial cloud signatures in the $\Delta$-variance analysis. Furthermore, $^{12}$CO is optically thick and thus we always find slightly steeper $\alpha$ slopes for the $^{12}$CO intensity tracers compared to the optically thin $^{13}$CO intensity tracers.

\subsection{Variation of the abundance of $^{13}$CO}
\label{subsec:ab13CO}

We also analyze the variation of the abundance of $^{13}$CO, because our simulations only follow the chemistry of $^{12}$CO, but not of its isotope $^{13}$CO. Therefore, we produce a set of $^{13}$CO emission maps, using a spatially varying $R_{12/13}$ generated following the prescription given in Section \ref{subsec:RADMC} and in \citet{SzucsEtAl2014}. After applying the radiative transfer post-processing to these maps, we compute the maps of the CV and the integrated intensity and compare the $\Delta$-variance spectra with each other. The results agree with those as described in \citeauthor{BertramEtAl2015a}~(2015a) for the centroid velocity increment structure functions. Although we encounter slight variations on small scales for both the CV and the integrated intensity maps, the spectra of the $\Delta$-variance do not change significantly as we vary $R_{12/13}$. Thus, the use of a constant $^{12}$CO/$^{13}$CO ratio is sufficient for obtaining proper $\Delta$-variance spectra for $^{13}$CO.

\section{Discussion}
\label{sec:discussion}

\subsection{Interpreting the $\Delta$-variance spectra computed using the CV maps}
\label{subsec:spectra}

Regarding the spectra computed on the CV maps in Fig. \ref{fig:DVspectra}, we find a good correlation between the total density and the H$_2$ density in the n300 model. This is because $\sim98$\% of the initial atomic hydrogen is in molecular form at this time. Thus, the spectra of the $\Delta$-variance for the CV maps are primarily dominated by the H$_2$ mass in this case. Vice versa, if we analyze the CV spectra of the n30 model, we find a larger discrepancy between the total density and the H$_2$ density, which is because only $\sim61$\% of the initial atomic hydrogen is in molecular form at this time. The n100 model is an intermediate case, with this value being $\sim78$\%. Furthermore, we obtain a good correlation between the various CO tracer components. This indicates, that the turbulence statistics are similar for all three cases and that the impact of the radiative transfer post-processing on the $\Delta$-variance analysis is limited. This is similar to the conclusion presented in \citeauthor{BertramEtAl2015a}~(2015a) based on the analysis of CV increment structure functions.

In general, we find that the slopes $\alpha$ for the total and H$_2$ density models are significantly steeper than the slopes for the different CO tracer models (see the values in Table \ref{tab:DVslopes}). This indicates that these components have a higher relative amount of structures on larger scales compared to the different CO tracers. Furthermore, as shown by \citet{GloverEtAl2010} and \citeauthor{ShettyEtAl2011b}~(2011a,b), CO is primarily a good tracer of dense and very compact regions in a cloud. Thus, this leads to less turbulent velocity structures on larger scales and hence to flatter slopes compared to those of the total and H$_2$ density. However, we find that the different slopes $\alpha$ of the various CO tracers underestimate the slopes of the total and H$_2$ density by a factor of $\sim1.5-3.0$ (see Table \ref{tab:DVslopes}).

Furthermore, we caution the reader that it is difficult to infer a clear dependence of the slopes on the mean ISM density. It is likely that the statistical measures we derive from our numerical simulations are also sensitive to the specific realization of the turbulent velocity field. Since we are studying flows which are driven on large scales, variance effects can become important, and the statistical properties depend on the random orientation of the turbulent modes as well \citep[see, e.g.][]{Klessen2000,KlessenEtAl2000}. Hence, in order to obtain slopes that properly converge with the density, we speculate that a large number of simulations with various turbulent realizations would be needed in order to calculate reliable average values \citep[see, e.g. the PCA analysis of the statistics using different realizations of the turbulent velocity field in][]{BertramEtAl2014}. This is prohibited by the high computational costs of the individual simulations, and hence we only focus on one example in this paper, which is enough to illustrate basic trends of the $\Delta$-variance statistics.

We also note that a direct comparison of our 2D $\Delta$-variance statistics to 3D turbulence models is difficult, since all our 2D maps are a complex convolution of the 3D density field (or the brightness temperatures) with the 3D velocity field. Various different physical processes can influence the CV statistics, as shown by previous studies. For example, \citet{LazarianEtAl2004}, \citeauthor{BurkhartEtAl2013a}~(2013a) and \citet{BurkhartEtAl2014} analyzed the impact of the sonic Mach number on the CV statistics, finding that it can significantly alter the results. Moreover, \citet{LazarianAndEsquivel2003}, \citet{OssenkopfEtAl2006}, \citet{EsquivelEtAl2007}, \citet{Hily-BlantEtAl2008} and \citet{FederrathEtAl2010} studied the effects of the turbulent driving as well as temperature and density fluctuations on the CV statistics, also finding significant differences in the statistics and in the inferred CV slopes. Nevertheless, although a direct comparison of the 2D statistical quantities to 3D measures is complicated, we can safely use the 2D $\Delta$-variance analysis in order to work out statistical trends measured in the spectra of the individual models.

\subsection{Interpreting the $\Delta$-variance spectra computed on the intensity and column density maps}
\label{subsec:WNspectra}

In Section \ref{subsec:slopesDV} and \ref{subsec:DVWN} we already justified the correlation between the total and the H$_2$ density runs in all our density models and compared the different slopes $\alpha$ from Table \ref{tab:DVWNslopes} with each other. Furthermore, we established that the observed CO gas distribution in the low-density n30 model does not reflect the spatial distribution of molecular hydrogen very well. This is because CO is mainly located in the dense filaments, while H$_2$ is more space-filling and distributed over the total cloud, owing to its greater ability to resist photodissociation. Consequently, the $\Delta$-variance spectra of the CO tracers peak at a scale corresponding to the width of these structures. Hence, if we apply the $\Delta$-variance analysis to the maps of integrated intensity or column density, we can obtain important information about the turbulently created high-density peaks within the MC \citep[see also the discussion in][]{OssenkopfEtAl2001}. However, we also note that the physical connection of our slopes $\alpha$ from Table \ref{tab:DVWNslopes} to observational measurements is complicated, since observations typically probe smaller spatial scales, which our simulations are not sensitive to due to the limited numerical resolution.

If we compare the CO tracers with the total and H$_2$ density in the n300 model, we find that the carbon monoxide better reproduces the statistical trends of molecular hydrogen in this high-density run compared to the low-density n30 model. In the n300 run, we generally obtain $\alpha > 0$ for all chemical components, i.e. we find more cloud structures on larger spatial scales. In contrast to the low-density n30 model, this is because carbon monoxide is not only confined to small dense filaments, but instead is distributed over the whole molecular cloud. In particular, a significant amount of CO gas can also be observed between the numerous dense cloud regions. Consequently, we see the $\Delta$-variance spectra peaking at the largest scales, consistent with observational efforts and previous work on this field \citep[see, e.g.][]{StutzkiEtAl1998,OssenkopfEtAl2001,BenschEtAl2001,SunEtAl2006,SchneiderEtAl2011,RusseilEtAl2013,AlvesdeOliveiraEtAl2014,EliaEtAl2014}.

However, regarding the general trend of the slopes $\alpha$ from the n30 to the n300 via the n100 model, we find that the values of $\alpha$ for the CO tracers change sign above a critical density of $\sim100\,$cm$^{-3}$. This is about the number density at which all values $\alpha$ become positive. Thus, we conclude that carbon monoxide traces our total cloud structure well only if the average cloud density lies significantly above a critical threshold of $\sim100\,$cm$^{-3}$. If the mean density in the cloud is significantly smaller than this limit, the observable CO gas does not properly trace the statistical properties of the H$_2$ gas in the cloud. We speculate that one may also see a similar switch from CO tracing all of the structures to only the dense cores and filaments if one increases the incident field strength or decreases the metallicity, as in both cases, this makes it much easier to photodissociate the diffuse CO (see, e.g. \citeauthor{GloverAndClark2012}~2012; \citeauthor{BertramEtAl2015b}~2015b, submitted). On the contrary, if the mean density in the cloud is too high, CO becomes optically thick and so we suspect that there should also exist an upper density limit above which CO does not properly trace structures of the cloud anymore. We leave such an analysis for further investigations.

\subsection{Comparison of the $\Delta$-variance to other statistical tools applied to our simulation data}
\label{subsec:tools}

We have previously applied several other statistical methods to the same set of simulation data in order to study the structural behavior of MCs. For example, \citet{BertramEtAl2014} applied Principal Component Analysis (PCA) to the same data used in this paper, while \citeauthor{BertramEtAl2015a}~(2015a) carried out a similar study using centroid velocity increment structure functions (CVISF) and Fourier spectra.

Comparing the results of the different methods with each other, we generally find consistent results between the different statistical analysis methods. For example, the slopes of the CVISF for the total density and H$_2$ density models are also significantly steeper compared to the various CO tracers, which is in good agreement to the results of the $\Delta$-variance analysis in this paper. The same holds for the relative scaling of the different CO models, which is the same in the analysis of the CVISF and the $\Delta$-variance. Interestingly, we also find a close relation between the PCA structure analysis and the spectra of the $\Delta$-variance for the CO tracers. The PCA method does not find any structures on larger scales for the CO tracers in the n30 model as well as for some single CO tracers in the n100 model. This result can also be reproduced in the different CO spectra of the $\Delta$-variance analysis, i.e. where the gradient of the individual spectra become negative at a characteristic scale of $\sim0.4\,$pc, as described in Sections \ref{subsec:DVWN}. This is the situation where small clumpy structures of CO gas dominate the overall composition of carbon monoxide in the MC. In this case, the $\Delta$-variance shows that the most dominant CO structures in the simulation domain can be found at small scales, leading to completely missing principal components on scales larger than $\sim1\,$pc, as presented in Fig. 3 in \citet{BertramEtAl2014}. Hence, applying the $\Delta$-variance analysis to maps of integrated intensities or column densities gives an idea about characteristic spatial scales in the cloud of interest. The $\Delta$-variance thus provides a good statistical tool in order to study the relative gas distribution on various cloud scales.

These results indicate that all these various statistical methods are connected to each other and that characteristic structural properties of the MC should be traced by each of them individually. Hence, all various statistical methods have proven to yield reasonable results for the structure analysis of MCs. Nevertheless, the advantage of the $\Delta$-variance method is that it is fast and easy to implement, while the computation of structure functions and principal components is more expensive. Thus, the $\Delta$-variance analysis provides a useful and adequate tool in order to quickly study the internal structure of a cloud. However, the advantage of the structure function analysis is that their results can be easily compared to theoretical models of turbulence.

\subsection{Previous studies about $\Delta$-variances computed using CV maps}
\label{subsec:previous1}

Several studies tried to compute the $\Delta$-variance spectra of CV maps in the past and estimated turbulent slope values $\alpha$ from observational measurements. For example, \citet{OssenkopfEtAl2008b} used optically thick $^{12}$CO ($J=1 \rightarrow 0$) maps of the Polaris Flare to compute centroid velocity maps and to estimate the slope $\alpha$. The Polaris flare is an archetype of a cloud midway between the diffuse and the molecular phases \citep{HeithausenAndThaddeus1990,MeyerdierksAndHeithausen1996,FalgaroneEtAl1998,AndreEtAl2010,Miville-DeschenessEtAl2010}. It is supposed to have a low average density and thus can be compared to our n30 low-density model. Adopting the same $^{12}$CO tracer, we find a value of $\alpha = 0.85 \pm 0.02$, in agreement with the estimate of $\alpha \approx 0.9$ found by \citet{OssenkopfEtAl2008b}.

Moreover, \citet{OssenkopfAndMacLow2002} computed the slope $\alpha$ for hydrodynamic supersonic simulations driven at different wavenumbers. They find one power-law range for all models and obtain $\alpha \approx 1.0$, which fits into our range of power-law slopes $\alpha$ from $0.8-1.3$ for the total density models. Furthermore, \citet{FederrathEtAl2009} computed $\Delta$-variance slopes for numerical simulations with both solenoidal and compressive forcing, finding a similar range of $\alpha$ values from $0.8-1.4$ for the turbulent velocity field. However, these simulations only use supersonic isothermal turbulence, while our runs include more complex physics, e.g. a chemical network, heating and cooling, various initial number densities or the coupling to the radiation field. Thus, if we compare the total variations of slopes $\alpha$ in their and in our models, we find that the influence of the different forcing methods on the slopes $\alpha$ in their simulations is large, while the impact of our additional physical effects (varying density, optical depth effects, etc.) on the slope values remains comparatively small.

\subsection{Previous studies about $\Delta$-variances computed using intensity and column density maps}
\label{subsec:previous2}

As well as computing $\Delta$-variance spectra for maps of centroid velocities, several studies also applied the $\Delta$-variance to maps of integrated intensity or column density. For example, \citet{BenschEtAl2001} computed $\Delta$-variance spectra on maps of velocity-integrated intensity for $^{12}$CO and $^{13}$CO ($J=1 \rightarrow 0$) for various MCs in the Galaxy. \citet{StutzkiEtAl1998} analyzed the $\Delta$-variance of an observed $^{12}$CO ($J=1 \rightarrow 0$) image of the Polaris flare as a whole, while \citet{OssenkopfEtAl1998} studied the intensity map of one of its subclouds, MCLD 123.5+24.9.

Comparing all the various $\Delta$-variance spectra in \citet{BenschEtAl2001} or \citet{StutzkiEtAl1998} with each other, we find a value $\alpha > 0$ for each of them in the given fitting range, i.e. the spectra increase with increasing spatial lag. At first sight, this is in contradiction to the results that we obtain in this study, where we measure $\alpha < 0$ for the various CO tracers in the n30 and n100 model, given our fitting range. However, this is primarily due to the fact that our $\Delta$-variance spectra peak at roughly the scale of small carbon monoxide clumps in the low-density clouds (see Section \ref{subsec:WNspectra}). Accordingly, we find more cloud structures on the scales of those localized CO structures, leading to negative $\alpha$ slopes in Figure \ref{fig:DVWNspectra}. A similar effect can be seen in the $\Delta$-variance analysis of $^{13}$CO ($J=1 \rightarrow 0$) maps of the outer Galaxy shown in Figure 3 of \citet{StutzkiEtAl1998}. On small scales, a positive power-law slope is recovered, but on large scales $\alpha$ becomes negative as one starts to smooth on scales larger than the typical size of the observed MCs. The same situation is also illustrated in the $\Delta$-variance spectra in \citet{SchneiderEtAl2011} for the $^{13}$CO ($J=1 \rightarrow 0$) molecular line survey of Cygnus X as well as in a $^{13}$CO ($J=2 \rightarrow 1$) study of the Perseus cloud in \citet{SunEtAl2006}, which peak at intermediate cloud scales. However, a $\Delta$-variance analysis of the small sub-regions NGC1333 and L1455 in the Perseus molecular cloud complex shows that $\alpha$ is always positive \citep[see Fig. 6 in][]{SunEtAl2006}, which is in agreement with our finding that a positive slope $\alpha$ might be related to a $\Delta$-variance analysis at lower spatial cloud scales, probing only very localized CO structures. Moreover, \citet{AlvesdeOliveiraEtAl2014} analysed the Chamaeleon molecular cloud complex using data from the Herschel Gould Belt Survey. This complex encompasses three MCs with different star formation histories. One of them is a quiescent cloud, which should be best comparable to our numerical simulations, since we neglect the effects of self-gravity. The $\Delta$-variance analysis of this MC also shows a clear break at mid to large scales \citep[see cloud Cha III in Fig. 5 in][]{AlvesdeOliveiraEtAl2014}, also displaying a characteristic spatial scale in the column density structure.

In our high-density n300 run, the CO gas is distributed over the whole MC and not only confined to small dense fragments. In this case, we also find cloud structures on larger spatial scales, leading to positive $\alpha$ slopes in Figure \ref{fig:DVWNspectra}.

\subsection{Model limitations}
\label{subsec:limitations}

Since we are running numerical simulations, we have to keep in mind that our runs are subject to various physical simplifications. In the current analysis, we focus on the impact of the chemistry and the opacity of the gas on our results, thus neglecting other physical processes that could bias our analysis. In first instance, we do not account for self-gravity and thus also do not model star formation or stellar feedback. More specifically we do not account for stellar radiation, SN feedback and other physical processes. We also do not include any large-scale dynamics, e.g. spiral arms or galactic rotation, although these are unlikely to be important on the 20\,pc scale studied here. Nevertheless, we can infer useful information about how the chemical composition of the gas and the opacity affect the $\Delta$-variance analysis. We also note that our results depend only weakly on the resolution and that this concerns mostly CO as a tracer molecule, as we show in Appendix \ref{app:resolution}. For future investigations, we want to analyze simulations that span a wider range of physical parameters, e.g. with different levels of the external radiation field, varying metallicities or additional physical processes, in order to find out how they affect the statistics.

\section{Summary and Conclusions}
\label{sec:summary}

We analyzed $\Delta$-variance spectra of MCs with time-dependent chemistry and radiative transfer post-processing for models of different initial number densities and chemical components: the total number density, H$_2$ and CO density (each without radiative transfer) as well as $^{12}$CO ($J=1 \rightarrow 0$) and $^{13}$CO ($J=1 \rightarrow 0$) intensity (both with radiative transfer). In each case, we computed $\Delta$-variance spectra for maps of centroid velocities (CV), integrated intensities and column densities and analyzed the structural behavior of MCs in numerical simulations. We report the following findings:
\begin{itemize}
\item We compute $\Delta$-variance spectra of maps of centroid velocities and fit a power-law $\sigma_\Delta^2 (\ell) \propto \ell^{\alpha}$, in order to characterize the properties of the turbulent hierarchy in the MCs. This power-law can be translated into a linewidth-size relation, i.e. $\sigma_\Delta (\ell) \propto \ell^{\gamma}$ with $\gamma = \alpha/2$, readily comparable to slopes that can be derived from spectral observations. We find the slopes $\alpha$ of both the total and H$_2$ density models to be significantly steeper than the slopes of the different CO tracers, which underestimate the former by a factor of $\sim1.5-3.0$ (see Section \ref{subsec:slopesDV}).
\item The slopes $\alpha$ derived from the CV maps range from 0.8 to 1.3 for the total and H$_2$ density, while $\alpha$ for the various CO tracers range from 0.3 to 0.8 (see Section \ref{subsec:slopesDV}). However, we also note that the specific choice of the fitting range might cause further variations of the slopes by $\pm0.1$.
\item Although we find slight variations between the different slopes $\alpha$ for our various CO models obtained from the CV maps, the impact of the optical depth effects on the spectra computed on maps of centroid velocities remains limited (see Table \ref{tab:DVslopes}).
\item This is different in the case of the integrated intensity and column density. The $\Delta$-variance computed from these maps is strongly affected by optical depth effects. The CO tracers exhibit a very different spatial scaling behavior compared to the total and H$_2$ density models (see Table \ref{tab:DVWNslopes} and Figure \ref{fig:DVWNspectra}).
\item We report a critical number density threshold of $\sim100\,$cm$^{-3}$ at which the spectral slopes $\alpha$ of the CO tracers switch sign for the $\Delta$-variance of integrated intensity and column density maps. We conclude that carbon monoxide traces the total cloud structure well only if the average cloud density lies significantly above this threshold. If the mean density in the cloud is significantly smaller than this limit, the observable CO gas does not properly trace the statistical properties of the H$_2$ gas in the cloud (see Section \ref{subsec:DVWN}).
\item The $\Delta$-variance spectra computed on maps of integrated intensity and column density provide a useful statistical measure in order to infer important information about the distribution of gas within a cloud. We also argue that peaks in the $\Delta$-variance spectra correspond to characteristic scales of the morphological structure of the system (see Section \ref{subsec:DVWN}).
\item Our findings are consistent with previous $\Delta$-variance studies using CO line observations or measurements of the continuum (see Section \ref{subsec:previous1} and \ref{subsec:previous2}).
\end{itemize}

\section*{Acknowledgements}

We thank Volker Ossenkopf for informative discussions about the usage of the $\Delta$-variance and for providing his IDL routines in order to compute the $\Delta$-variance spectra. We also thank Lukas Konstandin for stimulating discussions about the theory of turbulence as well as the referee for a very constructive and detailed report, which helped to improve the paper. EB, SCOG and RSK acknowledge support from the Deutsche Forschungsgemeinschaft (DFG) via the SFB 881 (sub-projects B1, B2, B5 and B8) ``The Milky Way System'', and the SPP (priority program) 1573, ``Physics of the ISM''. Furthermore, EB acknowledges financial support from the Konrad-Adenauer-Stiftung (KAS) via their ``Promotionsf\"orderung''. Some of the simulations presented in this paper were performed using the Ranger cluster at the Texas Advanced Computing Center, using time allocated as part of Teragrid project TG-MCA99S024. Additional simulations were performed on the \textit{kolob} cluster at the University of Heidelberg, which is funded in part by the DFG via Emmy-Noether grant BA 3706, and via a Frontier grant of Heidelberg University, sponsored by the German Excellence Initiative as well as the Baden-W\"urttemberg Foundation. RSK acknowledges support from the European Research Council under the European Community's Seventh Framework Programme (FP7/2007-2013) via the ERC Advanced Grant "STARLIGHT: Formation of the First Stars" (project number 339177).

\begin{appendix}

\section{Resolution study}
\label{app:resolution}

\begin{figure}
\centerline{
\includegraphics[height=0.73\linewidth,width=1.0\linewidth]{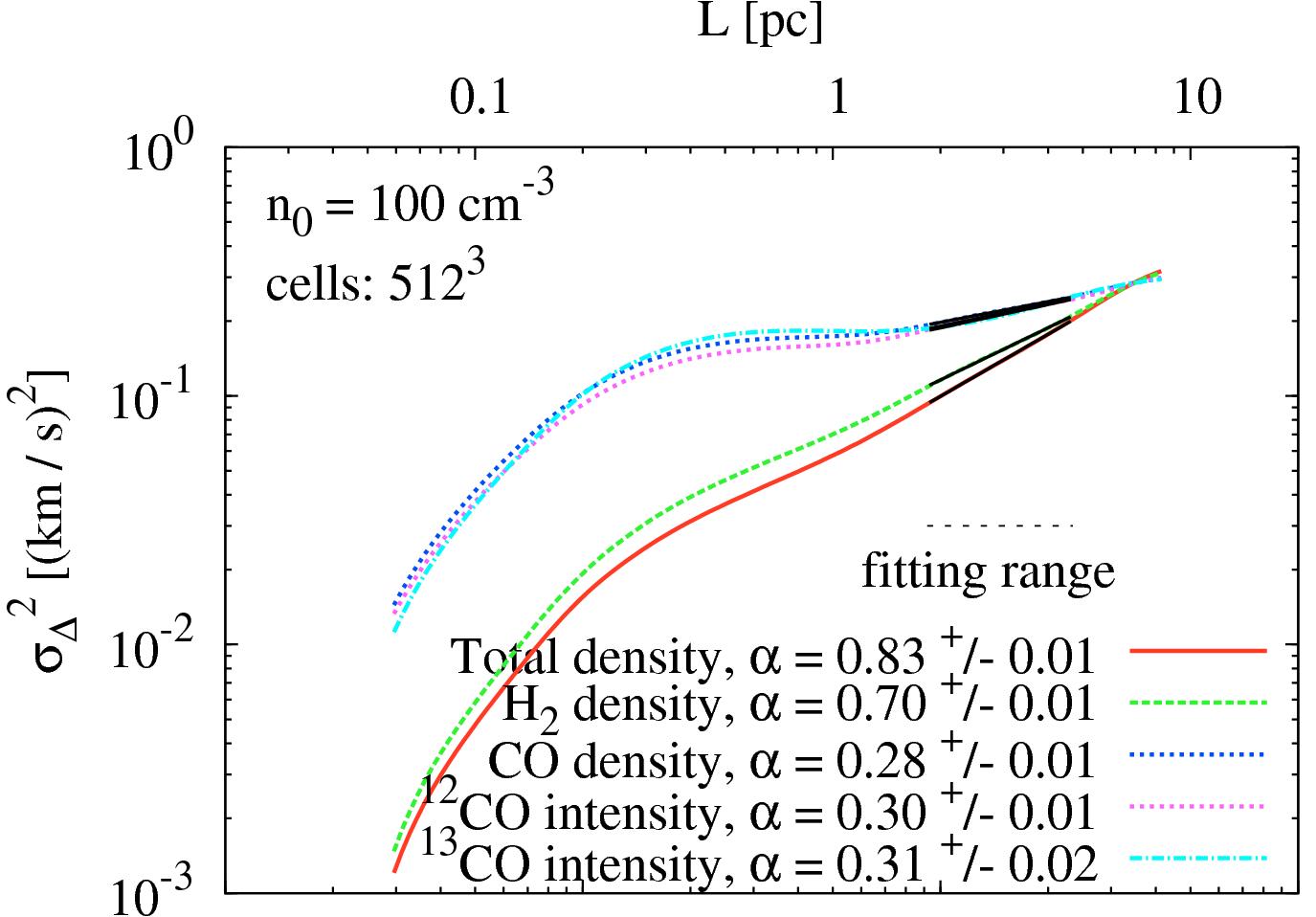} \\
}
\centerline{
\includegraphics[height=0.75\linewidth,width=1.0\linewidth]{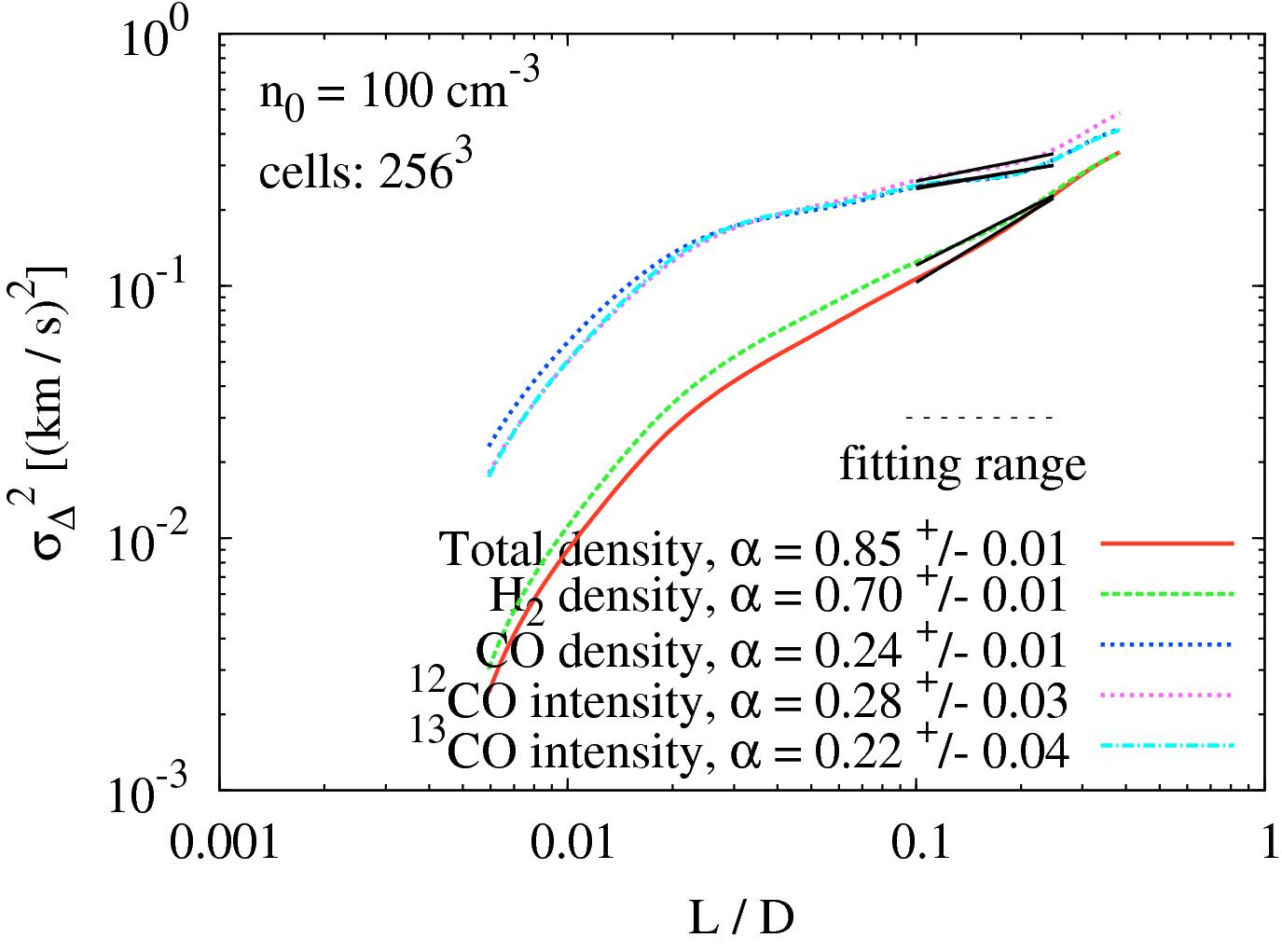} \\
}
\caption{Same as Fig. \ref{fig:DVspectra}, but with runs of different resolutions with $512^3$ and $256^3$ grid cells with a fixed initial number density of $n_0 = 100\,$cm$^{-3}$, computed for the centroid velocity (CV) maps. We find similar $\alpha$ values for the total density and the H$_2$ density models, while the values $\alpha$ differ by up to $\sim20-30\%$ between the different resolution models for the various CO tracers.}
\label{fig:DVresolution}
\end{figure}

We study the influence of the numerical resolution on the results of the $\Delta$-variance. Therefore, we have performed runs with resolutions of $512^3$ and $256^3$ grid cells and evaluate the spectra and slopes for the CV maps for all chemical components for a fixed initial number density of $n_0 = 100\,$cm$^{-3}$. The results and their interpretation is the same for all other density models. Fig. \ref{fig:DVresolution} shows spectra of the $\Delta$-variance with the corresponding slopes $\alpha$, in analogy to Fig. \ref{fig:DVspectra}. The fitting range for the $256^3$ model is downscaled by a factor of 2 compared to the $512^3$ model, i.e. we fit from 25 to 64 cells in the spatial domain. We find similar $\alpha$ values for the total density and the H$_2$ density models. However, the slope values $\alpha$ differ by up to $\sim20-30\%$ between the different resolution models for the various CO tracers. This is because CO is mainly located in dense regions of the cloud \citep{BertramEtAl2014}, which can be resolved more accurately at a higher resolution, leading to significant differences between the two resolution models. These results agree with the results in the resolution study of the structure function analysis presented in \citeauthor{BertramEtAl2015a}~(2015a). However, the variations caused by the effect of resolution are rather small, since $\alpha$ also strongly depends on the specific choice of the fitting range, which might also cause slope variations by about $\pm0.1$. Nevertheless, we find a similar relative scaling behavior in the two resolution models between the spectra of the different chemical components.

\section{Comparison of spectra with different filter functions}
\label{app:filters}

\begin{figure}
\centerline{
\includegraphics[height=0.80\linewidth,width=1.0\linewidth]{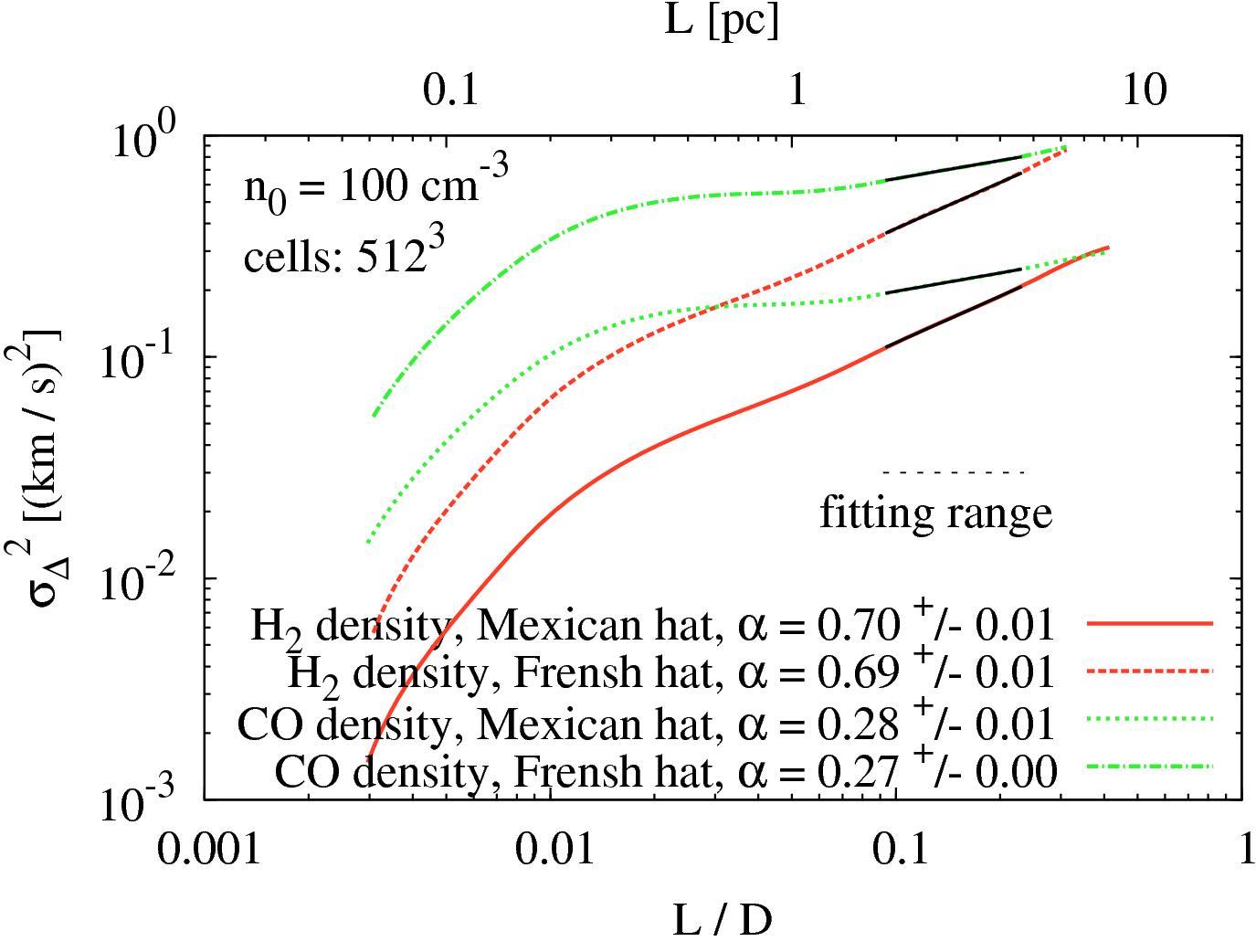} \\
}
\caption{Same as Fig. \ref{fig:DVspectra}, but with the $\Delta$-variance spectra for the H$_2$ and CO density models for a fixed initial number density of $n_0 = 100\,$cm$^{-3}$ and a resolution of $512^3$ grid cells, computed for the centroid velocity (CV) maps. In order to analyze the impact of the filter function and its diameter ratio on our results, we compute the H$_2$ and CO density spectra using both a Mexican hat with a diameter ratio of 1.5 and a French hat with a diameter ratio of 3.0. We do not find any significant differences in the slopes $\alpha$ within the fitting errors if we use a another filter function for the $\Delta$-variance analysis combined with a different diameter ratio.}
\label{fig:DVfilters}
\end{figure}

We also study the influence of the filter function and the choice of the specific diameter ratio on our $\Delta$-variance spectra. Therefore, as an example, Fig. \ref{fig:DVfilters} shows the $\Delta$-variance spectra for the H$_2$ and CO density models for a fixed initial number density of $n_0 = 100\,$cm$^{-3}$ and a resolution of $512^3$ grid cells, computed on maps of centroid velocities. For each model, we evaluate the spectra using different filter functions and diameter ratios. In particular, we compute the $\Delta$-variance spectra using a Mexican hat with a diameter ratio of 1.5 as well as a French hat with a diameter ratio of 3.0. In analogy to Section \ref{subsec:slopesDV}, we fit a power-law within a given fitting range to the spectra and compare the slopes with each other, which are shown in Fig. \ref{fig:DVfilters}. Thereby, we do not find any significant differences between the slopes derived from spectra with various filter functions and diameter ratios within the fitting errors for our models. The individual normalizations of the spectra are caused by the variable choice of the diameter ratio, affecting the computation of the $\Delta$-variance analysis \citep[see, e.g.][]{OssenkopfEtAl2008a}. However, the shape of the individual spectra for one chemical model over various spatial scales is also approximately conserved.

\bibliographystyle{mn2e}
\bibliography{lit/literature}

\end{appendix}
\end{document}